\documentclass[letterpaper]{article} % DO NOT CHANGE THIS
\usepackage{aaai24}  % DO NOT CHANGE THIS
\usepackage{times}  % DO NOT CHANGE THIS
\usepackage{helvet}  % DO NOT CHANGE THIS
\usepackage{courier}  % DO NOT CHANGE THIS
\usepackage[hyphens]{url}  % DO NOT CHANGE THIS
\usepackage{graphicx} % DO NOT CHANGE THIS
\usepackage[dvipsnames]{xcolor}
\usepackage{natbib}  % DO NOT CHANGE THIS AND DO NOT ADD ANY OPTIONS TO IT
\usepackage{caption} % DO NOT CHANGE THIS AND DO NOT ADD ANY OPTIONS TO IT
\usepackage{algorithm}
\usepackage{algorithmic}
\usepackage{enumitem}
\setlist{nolistsep}

\usepackage{newfloat}
\usepackage{listings}
\DeclareCaptionStyle{ruled}{labelfont=normalfont,labelsep=colon,strut=off} % DO NOT CHANGE THIS
\floatstyle{ruled}
\newfloat{listing}{tb}{lst}{}
\floatname{listing}{Listing}
\newcommand{\ethics}[1]{\textcolor{blue}{#1}}

\title{Supporters and Skeptics: LLM-based Analysis of Engagement with Mental Health (Mis)Information Content on Video-sharing Platforms}
\author {
    Viet Cuong Nguyen\textsuperscript{\rm 1},
    Mini Jain\textsuperscript{\rm 1},
    Abhijat Chauhan\textsuperscript{\rm 1},
    Heather Jaime Soled\textsuperscript{\rm 2},
    Santiago Alvarez Lesmes\textsuperscript{\rm 3},
    Zihang Li\textsuperscript{{\rm 4}}
    Michael L. Birnbaum\textsuperscript{\rm 3},
    Sunny X. Tang\textsuperscript{\rm 3}
    Srijan Kumar\textsuperscript{\rm 1}
    Munmun De Choudhury\textsuperscript{\rm 1}
}
\affiliations {
    \textsuperscript{\rm 1}Georgia Institute of Technology\\
    \textsuperscript{\rm 2}Rowan University\\
    \textsuperscript{\rm 3}Zucker Hillside Hospital, Psychiatry Research\\
    \textsuperscript{\rm 4}Hofstra University\\
    johnny.nguyen@gatech.edu
}

\newcommand{\red}[1]{\textcolor{teal}{#1}}

\usepackage{bibentry}
\begin{document}

\maketitle

\frenchspacing
\begin{abstract}
Over one in five adults in the US lives with a mental illness. In the face of a shortage of mental health professionals and offline resources, online short-form video content has grown to serve as a crucial conduit for disseminating mental health help and resources. However, the ease of content creation and access also contributes to the spread of misinformation, posing risks to accurate diagnosis and treatment. Detecting and understanding engagement with such content is crucial to mitigating their harmful effects on public health. We perform the first quantitative study of the phenomenon using YouTube Shorts and Bitchute as the sites of study. We contribute MentalMisinfo, a novel labeled mental health misinformation (MHMisinfo) dataset of 739 videos (639 from Youtube and 100 from Bitchute) and 135372 comments in total, using an expert-driven annotation schema. We first found that few-shot in-context learning with large language models (LLMs) are effective in detecting MHMisinfo videos. Next, we discover distinct and potentially alarming linguistic patterns in how audiences engage with MHMisinfo videos through commentary on both video-sharing platforms. 
Across the two platforms, comments could  exacerbate prevailing stigma with some groups showing heightened susceptibility to and alignment with MHMisinfo. We discuss technical and public health-driven adaptive solutions to tackling the ``epidemic'' of mental health misinformation online.

% We aim to answer three key questions: (1) Can we effectively identify mental health misinformation using Large Language Models (LLMs)? (2) How do users engage with and react to misinformative mental health videos, when compared to non-misinformative mental health videos? (3) Are stigmatizing comments more likely with misinformation content? We contribute a novel mental health misinformation dataset \red{containing 639 videos and 134347 comments}, labeled using an expert-driven schema. In addition, we also showed that a few-shot in-context learning LLM-based approach can effectively and efficiently detect misinformative videos when comparing to gradient-based approaches trained on substantially more data. Finally, our LLM-enabled findings reveal distinguishing linguistic and sentiment factors of engagement to mental health misinformation. Our findings guide recommendations to address misinformation and improve mental health information quality on video-sharing platforms. \red{Overall, this research paves the way for future work on designing adaptive interventions against mental health misinformation and utilizing LLMs to understand online mental health discourse.}
\end{abstract}

\section{Introduction}
\begin{figure*}[t]
\centering
        \includegraphics[scale=0.15]{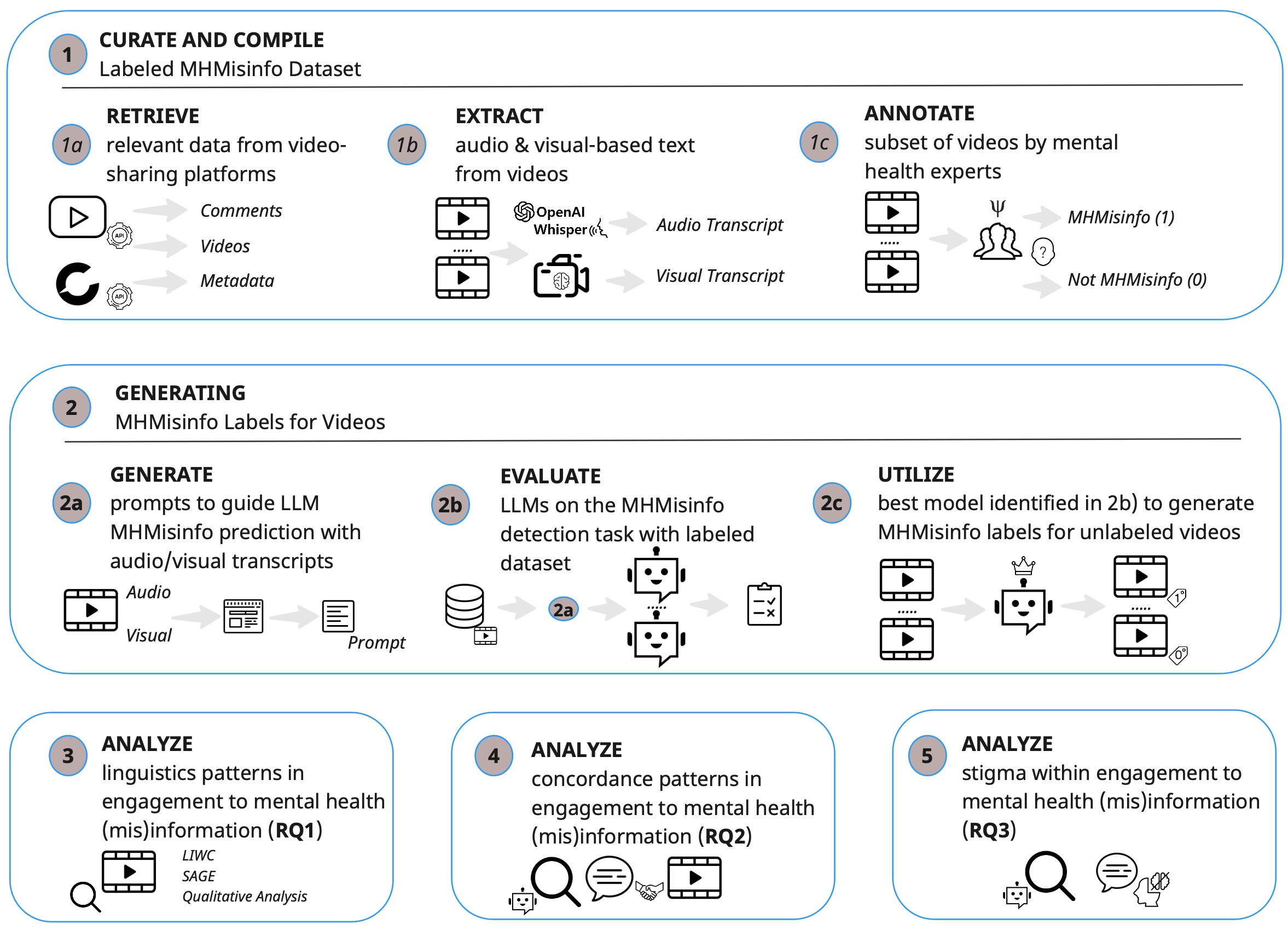}
    \caption{Diagram summarizing our research design}
    \label{fig:verticalcell} 
\end{figure*}

In an article in Time 
\footnote{\url{https://time.com/6307996/mental-health-content-red-flags-social-media/}}, Angela Haupt wrote, ``{\it The classic vision of therapy revolved around a person on a couch, supine, tapping into their deepest and darkest hopes and fears. A modern-day remix might look like this: a person still on a couch, but at home, scrolling through a constantly refreshing selection of mental-health content on social media platforms like TikTok and Instagram.}'' 

% Background
The global crisis of limited access to mental healthcare, aggravated by pervasive stigma, poses a significant threat to public health \cite{ Mental_Health_America_2023}. This crisis is further compounded by a critical shortage of mental health professionals, particularly in underserved areas designated as Mental Health Professional Shortage Areas (MPSAs) \cite{andrilla2018geographic}. Consequently, in recent years, people in distress have begun to seek health information online, as a way to fulfill their unmet mental health needs. 
Users engaging with mental health content on social media platforms like YouTube and Reddit often find a sense of community and belonging in the comments sections \cite{milton2023see}.

% Importance
However, the ease of access and dissemination of information on social media platforms also presents a significant challenge: the proliferation of misinformation. Misinformation about mental health is detrimental, as it can lead to inaccurate self-diagnosis, self-stigma, delayed help-seeking and treatment initiation and worsened symptoms\cite{starvaggi2023mental}. The stigma surrounding mental illness already serves as a barrier to seeking help, and misinformation can further exacerbate this issue by promoting harmful stereotypes and inaccurate information about diagnoses and treatment options \cite{corrigan2004stigma}. 

% Problem
Despite the potential for severe harm, research on mental health misinformation (MHMisinfo) and its impact on social media platforms remains comparatively scarce compared to other health topics \cite{starvaggi2023mental}. Underexploration of the phenomenon represents a significant barrier to mitigating the negative effects of MHMisinfo on public health. Understanding the presence of MHMisinfo content on social media and how users engage with such content is crucial for several reasons. First, it allows us to identify which misinformative narratives resonate most with the public, highlighting areas where education and intervention are paramount. In addition, given previous research on the role misinformation plays in reinforcing stigmatizing views against mental health, it is also important to explore the potential negative effects of online MHMisinfo through the presence of stigma within engagement to such content \cite{bizzotto2023buffering}.

Addressing the understudied area of online MHMisinfo content, we focus on the following research questions.
\begin{itemize} 
    \item \textbf{RQ1}: Do MHMisinfo and non-MHMisinfo videos differ in terms of their linguistic patterns of user engagement? % on MHMisinfo videos differ from that on non-MHMisinfo videos?
    \item \textbf{RQ2}: Do comments to MHMisinfo videos have comparable agreement rates with non-MHMisinfo videos?  
    \item \textbf{RQ3}: Are comments to MHMisinfo videos more likely to perpetuate mental health 
    stigma compared to non-MHMisinfo videos?
\end{itemize}

We study engagement in short-form mental health video content on two video-sharing platforms: YouTube (more specifically YouTube Shorts) and Bitchute. We chose YouTube as our main platform of study as it is the most popular video-sharing platform globally \cite{hussein2020measuring} and also a place where mental health information is abound$^1$. In addition, we also collect data from Bitchute for additional insights as it has been reported to have high rates of misinformation due to its alt-tech, low-moderation design \cite{trujillo2020bitchute}. We choose to focus on short-form videos, as this form of content is popular among mental health content creators and consumers alike due to their accessibility in both the creation and consumption of these videos~\cite{basch2022deconstructing}. 
Our contributions in this paper can be summarized as follows:

\begin{enumerate}
    \item We create a novel human-labeled mental health misinformation dataset, \texttt{MentalMisinfo}, containing 639 YouTube videos (13.14\% MHMisinfo) and 100 Bitchute videos (36\% MHMisinfo). Along with the videos, the dataset also includes 134347 YouTube comments (6.1\% in response to MHMisinfo videos) and 1025 Bitchute comments (23.7\% in response to MHMisinfo videos).
    \item We propose an expert-driven schema to label MHMisinfo videos accurately and rigorously based on three factors: Information on Interventions, Alignment with Medical Consensus, and Evidence-based Treatment.
    \item  We derive valuable linguistic insights on engagement to MHMisinfo content, demonstrating that certain groups (those using religious, male-oriented language) are likelier to be susceptible to MHMisinfo.
    \item Through an LLM-based approach, we found troubling trends regarding engagement on MHMisinfo content on both platforms, including higher rates of stigmatizing views against people with mental health conditions and substantial agreement rate to MHMisinfo content.
    \item We suggest potential adaptive interventions and content moderation strategies, grounded in both technical and public health approaches, against MHMisinfo.
\end{enumerate}

\vspace{0.05in} \noindent {\bf Content Warning and Disclaimer.} 
Consistent with best practices for working with public social media data, we report paraphrased quotes to ground some of our findings. We caution that some quotes may be distressing, stigmatizing, perpetuating stereotypes, or not backed by scientific evidence in mental health. Misinformative examples are provided only for illustrative purposes within the findings; these data should not be used for treatment or intervention.  

\section{Background and Related Works}
\subsection{Health (Mis)information on Social Media}

With the rise in LLM-based chatbots and the ever-increasing digitization of social interactions, the study of public interaction and rapid dissemination of health misinformation has increased in urgency and scope. Past research has shown that health misinformation tends to be prevalent in discussions of topics such as vaccinations, pandemics, medical treatments, eating disorders, and drugs and smoking \cite{suarez2021prevalence}. One of the gaps of knowledge when studying health misinformation is the impact of social media platforms on the dissemination of health misinformation as well as its impact on behaviors of users and the general public \cite{suarez2021prevalence}. Take the example of vaccine hesitancy, or the reaction towards social media content discouraging the use of vaccines \cite{van2022investigating, wawrzuta2021characteristics}. Van Kampen et al showed through a qualitative study involving two independent reviewers that TikTok videos spreading vaccination discouragement messages tended to originate from laypersons in contrast with health care professionals. As a result of this lack of access to accurate information about vaccines, the public resorts to ideas of freedom of choice, alternative health practices, and vaccine effectiveness skepticism\cite{wawrzuta2021characteristics}. Sentiment analysis, context/text analysis, social network analysis, and evaluation of the quality of content have been used to map the health misinformation and social media topography \cite{suarez2021prevalence}. Our work complements and builds on previous work in this area by examining mental health misinformation on social media, an understudied yet still important area of health misinformation. 

\subsection{Mental Health and Social Media}
There has been a breadth of research that explores the intersection between mental health and social media. Experiencing mental health issues is challenging for most people, and social media has been identified as a site where social support between those with mental health challenges can be both given to and gained from others \cite{hanley2019systematic}. In addition, to understand who seeks social support regarding mental health and why, researchers have also been interested in detecting and analyzing instances of self-disclosure (i.e. disclosing their mental health status despite the pervasive stigma) on social media \cite{de2014mental}. % For instance, De Choudhury et al. (2014) found that, through analyzing 20,411 mental health-related Reddit posts, that hat people use Reddit as a venue of self-expression regarding their experiences around their illness challenges, as well as the impact of those experiences on their work, life, and  relationships \cite{de2014mental}. 
Finally, there has been a great amount of interest from both computational social scientists and mental health professionals to investigate how user-generated activity and linguistic traces on social media could be used to infer their mental states \cite{eichstaedt2018facebook}. Previous research has primarily been focused on predicting depression and suicidal ideation through machine learning models trained on aggregated user-level Twitter and Reddit data, achieving substantial predictive performance~ \cite{chancellor2020methods}. Our work tackles a novel line of inquiry in this area of research, cognizant of social media's role as an ubiquitous source of health information, by examining the phenonemon of mental health misinformation on video-sharing platforms and distinguishing characteristics of engagement to such content. 

\subsection{LLMs for Social Media Analysis}
Social media is a massive repository of real-time data and analyzing this data effectively requires advanced tools capable of understanding complex language, context, and framing. LLMs offer potential in this domain, with their ability to process large amounts of text and extract meaningful insights. Therefore, LLMs have increasingly been used for various social media analysis tasks \cite{shen2023shaping}, especially in areas where availability of labeled data is rare by incorporating human insight into the model through prompting. LLMs are able to capture the patterns within the input prompt and are able to process it with their strong semantic understanding and reasoning abilities to adapt to diverse tasks at inference time. This is achieved through in-context learning, wherein the model is fed a series of labeled examples for the problem and then asked to apply that learning to unseen examples. Recent works have used in-context learning to imbibe few-shot or zero-shot abilities in LLMs such as GPT-3, LLaMA for tasks such as depression detection \cite{qin2023read} and multilingual misinformation detection \cite{pelrine2023towards}, outperforming existing gradient learning based techniques. Plaza-del-Arco et al. (2023) compared conventional transformer models against LLMs with zero shot prompting for hate speech detection and have found LLM-based models to be on par and sometimes better than conventional gradient learning models \cite{del2023respectful}. MentaLLaMa is the first open source fine-tuned LLM for mental health analysis tasks on social media data \cite{yang2023towards}. To our knowledge, our work is the first to explore the use of LLMs with in-context learning for detection and understanding of mental health misinformation on video-sharing platforms.

\section{The MentalMisinfo %: Mental Health Misinformation 
Dataset}
\begin{table*}[t]
\begin{tabular}{llllllll}
Platform & Category        & \# Videos & \# Creators & Av. Length (s) & Av. \#Views & Av. \#Likes & Av. \#Comments \\ \hline
YouTube & NMisinfo & 555  & 442 &     27.00           & 482,458         & 20909           & 287              \\
& Misinfo     & 84        & 77          &    34.81           & 60,537          & 3945            & 104  \\ \hline     
Bitchute & NMisinfo & 64       & 47         &     113.66           & 822         & 21           & 12              \\
& Misinfo     & 36        & 32          &    139.83           & 4815          & 19            & 6  \\ \hline   
\end{tabular}
\caption{Summary statistics for the labeled Misinfo and non-NMisinfo videos within \texttt{MentalMisinfo}}\label{tbl:summary_stat_videos}
\end{table*}

\subsection{Data Collection}
% \red{\subsubsection{YouTube}}
\paragraph{YouTube Shorts} We downloaded and gathered metadata for all public YouTube Shorts mental health videos using the official YouTube Data API and a search query with 22 relevant keywords, as detailed in Table \ref{tbl:dist_keywords_appendix}. These keywords, derived from prior social media and mental health research \cite{mittal2023moral} and re-validated by co-authors who are mental health professionals as well as researchers, 
capture salient mental health content on YouTube Shorts. While acknowledging the keyword set may not encompass the entire mental health discourse on YouTube, it effectively captures common conditions and phrases in mental health content on video-sharing platforms. All collected videos and comments are publicly accessible on YouTube without requiring authorization. We amassed relevant videos within a 12-month timeframe from October 25, 2022, to October 25, 2023, resulting in 33,686 videos.

\paragraph{Bitchute}
As Bitchute does not have a section dedicated to short-form videos like YouTube, we downloaded and gathered metadata for all Bitchute videos shorter than 5 minutes that matches one or more keywords from the set described in Table \ref{tbl:dist_keywords_appendix} through web scraping. Collecting all such Bitchute videos created between October 25, 2022, to October 25, 2023 yields 2485 videos.

\subsection{Data Filtering and Cleaning}
Focusing on English-language mental health videos and user reactions, we employed \textit{lingua}, a language detection library\footnote{\underline{\url{https://pypi.org/project/lingua-language-detector}}}. We chose the library due to its reported 99.2\% accuracy in identifying English sentences and its speed compared to similar libraries. We use it to exclude non-English videos based on their titles. Our filtering resulted in 19,010 English-language YouTube videos (2453 videos for Bitchute).
Given our emphasis on informational content related to mental health, we applied a category-based heuristic to remove YouTube videos using the specified keywords in non-medical contexts, such as those categorized by YouTube as ``Gaming," ``Music," ``Comedy," ``Sports," or ``Entertainment." Given the negligibility of such non-medical content on Bitchute in addition to the unavailability of topic metadata associated with Bitchute videos, we do not perform this data filtering step with collected Bitchute data. At the end of the data filtering process, we retained 16,785 YouTube videos (49.82\% of the collected videos) and 2453 Bitchute videos (98.7\% of the collected videos)
We then comprehensively gathered all comments for YouTube videos, including responses to replies to the videos above via the official API, totaling 569847 comments. For Bitchute videos, we follow the same procedure through web scraping with Selenium. which yields 13685 comments in total.  Notably, only 7,294 YouTube videos and 887 Bitchute videos (21.65\% for YouTube, 31.88\% for Bitchute) received at least one comment, with an average of 84.17 comments per YouTube video (10.68 for Bitchute videos) among this subset.
Table \ref{tbl:dist_keywords_appendix} provides a distribution of videos corresponding to each keyword collected for each platform along with exemplar video titles.

\subsection{Data Annotation Approach: Schema Development}
Without a previous peer-reviewed schema for identifying mental health (MH) misinformation in social media content, we devised our schema through an iterative process. Defining MH misinformation as ``Medically-relevant claims about mental health which are partially or fully false," we engaged co-authors with expertise in psychiatry, misinformation, and computational studies of social media in a brainstorming session. Here, we identified potential misinformative claims from a medical perspective. The first author employed affinity diagramming to group these claims into generalizable criteria. Each criterion was assessed using a 3-point Likert scale (-1, 0, 1). Five human annotators, including four co-authors, validate these criteria through individually annotated 10 randomly-selected videos. Subsequent discussions and criteria refinement among the annotators led to consensus on three main criteria for determining the informativeness of mental health-related videos, along with the annotation heuristic for each.
\begin{enumerate}
    \item \textbf{Information on Interventions:} Focus solely on the factuality of the information regarding treatments presented in the video (if any)
    \item \textbf{Alignment with Medical Consensus:} Similar to “Information on Interventions”, but for information regarding mental health topics that are non-treatment-related (e.g. symptoms, diagnosis) 
    \item \textbf{Evidence-based Treatment:} Focus solely on whether the author encourages the usage of evidence-based treatment (EBT) or not (e.g. clinically-tested medication, etc.) If the author encourages the usage of some EBTs and discourages others solely on an anecdotal basis, then a negative value should be assigned for this criterion.
\end{enumerate}

\begin{table}[t]
\centering
\small
\begin{tabular}{l@{}l@{}lll}
Platform\,\, & Category\,\,           & \# Comments & \# Commenters & Av. Length \\ \hline
YouTube\,\, & NMisinfo\,\, & 126555       & 109509         &   103                 \\
& Misinfo\,\,     & 7792        &    6727      &       111  \\
\hline
Bitchute\,\, & NMisinfo\,\, & 792       & 511         &   148                \\
& Misinfo\,\,     & 233        &    156      &       308  \\
\end{tabular}
\caption{Summary statistics for the comments associated with the labeled MHMisinfo and non-MHMisinfo videos, categorized by platform within \texttt{MentalMisinfo}} \label{tbl:summary_stat_comments}
\end{table}

\subsection{Gathering Human Annotations}
To create a labeled dataset, we recruited three expert annotators who are mental health professionals affiliated with two supervising co-authors' medical institutions. Given time and effort constraints, we annotated a subset of videos with the most comments. We randomly sampled 650 videos from the top 1000 most-commented YouTube videos and 100 from the top 1000 most-commented BitChute videos. Annotators assessed these videos for mental health misinformation (MHMisinfo) using a 3-point Likert scale for three MHMisinfo criteria.

Annotators underwent a training session, annotating 10 videos with the lead author, to reinforce the annotation schema. Videos were sent to annotators in batches of 100, with annotators labeling 20 videos per day to minimize fatigue. Annotators watched each video entirely on YouTube's native interface before annotation.

An annotator labeled a video as MHMisinfo (referred to as the cumulative MHMisinfo label) if they scored any of the three criteria negatively. A video is considered non-MHMisinfo only if all three criteria-level scores are non-negative. Any video cumulatively labeled by at least two out of three annotators as MHMisinfo is considered MHMisinfo for all subsequent analyses.

High inter-rater agreement was observed for both YouTube and BitChute data. For YouTube, Randolph's \(\kappa\) values for the three MHMisinfo criteria were: Information on Interventions: 0.902, Evidence-based Treatment: 0.858, Alignment with Medical Consensus: 0.727, and cumulative MHMisinfo: 0.807. For BitChute, the corresponding \(\kappa\) values were: Information on Interventions: 0.71, Evidence-based Treatment: 0.7, Alignment with Medical Consensus: 0.675, and cumulative MHMisinfo: 0.76.

A total of 11 YouTube videos (1.5\%) were not annotated due to being no longer publicly available at the time of annotation and were excluded. At the end of the annotation process, 555 YouTube videos and 64 BitChute videos were labeled non-MHMisinfo (86.8\% YouTube, 64\% BitChute), while 84 YouTube videos and 36 BitChute videos were labeled MHMisinfo (13.2\% YouTube, 36\% BitChute). Among the YouTube MHMisinfo videos, 55 (65.5\%) had low Alignment with Medical Consensus, 40 (47.6\%) had inaccurate Information on Interventions, and 11 (13.1\%) discouraged the usage of Evidence-based Treatment. For BitChute, the corresponding numbers were: 35 videos (35\%) had low Alignment with Medical Consensus, 15 (15\%) had inaccurate Information on Interventions, and 22 (22\%) discouraged the usage of Evidence-based Treatment. There were 7792 comments associated with labeled YouTube MHMisinfo videos (233 for BitChute videos) and 126555 comments associated with labeled YouTube non-MHMisinfo videos (792 for BitChute videos). 
\subsection{Dataset Processing}
To prepare the labeled video dataset for subsequent experiments, we extracted the audio transcript of all videos using OpenAI's Whisper automatic speech recognition (ASR) tool\footnote{\underline{\url{https://github.com/openai/whisper}}}. The tool has a reported 4\% word error rate on FLEURS, a popular ASR benchmark, and has been utilized in previous research for ASR on video data. We also extracted the video's on-screen text content using Google's Video Intelligence API\footnote{\underline{\url{https://cloud.google.com/video-intelligence/docs}}}. This text content will be referred to as the \textit{visual transcript} moving forward.

\section{Detection of Mental Health Misinformation}
\subsection{Methods}
To enable studying engagement to mental health misinformation (MHMisinfo) at scale, we create machine learning models to detect MHMisinfo videos trained using our \texttt{MentalMisinfo} dataset.
% Automatic detection of misinformative mental health-related videos is necessary for platform integrity and public health efforts \cite{wang2023health}. In addition to allowing for analysis of users' reactions to such videos, accurate identification of mental health misinformation also enables the deployment of targeted moderation efforts, which can include provision of verified mental health resources, search demotion, fact-checking labels, or even removal if necessary.  

\subsubsection{Problem Formulation}
We formulate the MHMisinfo video detection problem in this paper as a binary classification problem task. Given a video $D$, it can be classified as either MHMisinfo ($y = 1$, positive class) or non-MHMisinfo ($y = 0$, negative class). A video $D$ is expressed as a triple $(T, A, V)$, where $T$ is the title of the video, $A$ is the audio transcript of the video, and $V$ is the visual transcript of the video (i.e. the on-screen text aggregated across all video frames). While we acknowledge that other features can be extracted from video data, such as vocalic and visual features, we do not include them to simplify prompting for LLM-based classification. We aim to create a classifier $F: F(T;A;V) \longrightarrow y$, where $y \in \{0, 1\}$ is the ground-truth label of a video. We apply three different learning paradigms for learning this classifier $F$: gradient descent-based learning, anomaly detection, and in-context learning.
\subsubsection{In-context Learning}
In-context learning (ICL) is a transfer learning strategy that does not update the weights of the pre-trained LLM, unlike gradient learning strategies \cite{brown2020language}. ICL adapts a model to a task by conditioning it on a sequence of demonstrative ``examples". A demonstration typically refers to a pair $(x, y)$, where $x$ is the input and $y$ is the ground-truth label, which is then rendered using a natural language template. ICL thus feeds the model a prompt containing a sequence of such demonstrations, followed by the test input (also rendered in the template) \cite{min2022rethinking}. The language model is then expected to predict the label of this final data point by generating the most likely sequence of tokens following the input prompt. In-context learning has been shown to perform well in diverse tasks where labeled data is rare or difficult to gather and allows for flexibly incorporating human knowledge into large language models (LLMs) through updating the demonstrations and the prompt template \cite{min2022rethinking}. 

We apply ICL to detect MHMisinfo, using the following as base LLMs for comparative purposes. We choose these models due to their widespread usage and performance in LLM benchmarks, and for one model (\texttt{MentalLLaMa}), purported expertise in mental health analysis tasks \cite{yang2023towards, zhang2023benchmarking}.

\subsubsection{LLMs used:} 
\paragraph{Flan-T5.}
\texttt{Flan-T5} is a family of LLMs introduced in \cite{chung2022scaling}. For our experiment, we use the XL variant of \texttt{Flan-T5}, which contains about 3 billion parameters. %  \texttt{Flan-T5} is an enhanced version of the T5 architecture which aims to improve task generalization by finetuning the model on a variety of datasets converted into instructions \cite{chung2022scaling}.

\paragraph{LLaMa.}
%\srijan{dont give history of these models. tell which model size you use and why.}
\texttt{LLaMa} is a family of LLMs introduced in \cite{touvron2023llama}. For our experiment, we use the 7B and 13B variants of \texttt{LLaMa-2} % and the 8B variant of LLaMa-3. The second generation of LLaMa models, released in July 2023, includes 7 billion, 13 billion, and 70 billion parameter variants. LLaMa's weights, unlike that of many other LLMs, are open-source which allows for third-party efforts to fine-tune these models for task-specific domains.

\paragraph{MentalLLaMa.}
\texttt{MentalLLaMa}, introduced by Yang et al. (2023), is the first open-source LLM specifically built for mental health analysis tasks \cite{yang2023towards}. For our experiment, we use the 7B and 13B variants of \texttt{MentaLLaMa}.

\paragraph{Mistral.}
\texttt{Mistral} is a state-of-the-art open-source LLM by \cite{jiang2023mistral}. We use the \texttt{Mistral-v0.1} variant, which contains about 7 billion parameters.
% While only containing 7 billion parameters, it has been shown to out-perform LLaMa2-13B on 19 common LLM benchmarks involving commonsense reasoning, world knowledge, and reading comprehension and, for some benchmarks, even LLaMa2-34B while having substantially faster inference time.

\paragraph{Generative Pre-trained Transformer (GPT).}
% \srijan{dont give history of these models. tell which model size you use and why.}
Generative Pre-trained Transformer (\texttt{GPT}) is a family of large language models released by OpenAI \cite{brown2020language}. For our experiment, we utilize the gpt-3.5-turbo (henceforth referred to as \texttt{GPT3.5}) and gpt-4-0613 models (henceforth referred to as \texttt{GPT4}) models. % GPT models leverage the transformer architecture to pre-train on extensive datasets. This enables the generation of coherent and contextually relevant human-like text across a diverse range of tasks without any task-specific fine-tuning. Unlike all other models used in this paper, GPT models are not open-sourced, and can only be accessed through OpenAI's official API.

% \newline\newline

We train all of these models using 5-shot ICL, i.e. 5 demonstrations are used to adapt the models to the task. We select the 5-shot setting, as it is the most commonly-used setting in previous few-shot learning experiments \cite{parnami2022learning}. We use the same 5 shots, 3 MHMisinfo, and 2 non-MHMisinfo, to train all LLM-based models. The shots are randomly selected from expert-labeled YouTube and Bitchute videos as described in the previous section. The trained models are then evaluated on the remaining unlabeled data points. For all models using ICL, we design the input prompt template given in Table \ref{tbl:prompt_misinfo}. 
% \newline\newline

% \textit{You are an expert psychiatrist who has comprehensive knowledge of all mental health conditions and the misinformation surrounding them. You are helpful, so you will try your best to give accurate answers. Now, thoroughly look at the following examples which contain the audio transcription and rationale on whether a video contains misinformation regarding mental health or not.
% \newline \newline 
% Example: [Example 1 Data]. 
% \newline 
% Answer: [Answer]
% \newline 
% ......
% \newline 
% Example: [Example 5 Data]. 
% \newline 
% Answer: [Answer] 
% \newline \newline 
% Now, only using the video data provided below, answer whether the video contains mental health misinformation or not. You must include the word "yes" or "no" within your answer, then give your reasoning.
% \newline \newline 
% [Target Data]. 
% \newline 
% Answer:
% }
%\newline\newline

\begin{table}[!h]
\centering
\small 
\setlength{\tabcolsep}{1.0pt}
\begin{tabular}{l@{}lccccc}
\textbf{Model} & \textbf{Param} & \textbf{F1-} & \textbf{F1-} & \textbf{Prec} & \textbf{Rec} \\ 
& & \textbf{Macro} & \textbf{Pos} & &\\ \hline
\multicolumn{7}{|c|}{\textbf{Zero-shot}}                                                                                     \\ \hline
\texttt{Flan-T5}                                 & ~3B                                      & 0.605       & 0.301  & 0.623 & 0.594\\
 & & (0.493)       & (0.24)  & (0.54) & (0.521)        \\
\texttt{Mistral-0.1}                             & ~7B                      & 0.390                  & 0.196               & 0.489 & 0.470             \\
 & & (0.452)                 & (0.386)               & (0.465) & (0.463)         \\
\texttt{LLaMa-13B}                                & 13B                      & 0.436                  & 0.089               & 0.457 &  0.429           \\
 & & (0.39)        & (0)  & (0.32) & (0.5)          \\
\texttt{MentaLLaMa-13B}                          & 13B                      & 0.546                  & 0.505               & 0.565 & 0.569           \\
 & & (0.566)        & (0.527)  & (0.585) & (0.591)          \\
\texttt{GPT3.5}             & 175B                   & 0.514                      & 0.287                  & 0.550               & 0.604           \\
 & & (0.427)        & (0.095)  & (0.486) & (0.496)          \\
\texttt{GPT4}                                     & 8 x 220B                      & 0.707                  & 0.517               & 0.681 & 0.771            \\ 
 & & (0.669)        & (0.609)  & (0.669) & (0.683)          \\
 \hline
\multicolumn{7}{|c|}{\textbf{Few-shot with In-context Learning}}                                                             \\ \hline
\texttt{Flan-T5}                                  & ~3B                      & 0.614                  & 0.322               & 0.624  &  0.607\\
 & & (0.531)        & (0.339)  & (0.548) & (0.537         \\
\texttt{Mistral-0.1}                             & ~7B                      &   \textbf{0.626}                & 0.364              & 0.615 & 0.642             \\
 & & (0.613)        & (0.493)  & (0.616) & \textbf{(0.611)} &          \\
\texttt{LLaMa-7B}                                 & ~7B                      & 0.561                  & 0.217               & 0.58 & 0.554              \\
 & & (0.469)        & (0.15)  & \textbf{(0.703)} & (0.534)          \\
\texttt{MentalLLaMa-7B}                           & ~7B                      & 0.603                  & 0.328               & 0.594 & 0.619             \\
 & & (0.583)        & (0.5)  & (0.584) & (0.590)         \\
\texttt{LLaMa-13B}                                & ~13B                      & 0.476 & 0.023 & \textbf{0.684} & 0.505             \\
 & & (0.555)        & (0.511)  & (0.572) & (0.577)         \\
\texttt{MentaLLaMa-13B}                          & ~13B                      & 0.595                  & \textbf{0.373}               & 0.597 & \textbf{0.685} &              \\
 & & (0.546)        & \textbf{(0.5054)}  & (0.565) & (0.569)         \\
\texttt{GPT3.5}                                   & ~175B                      & 0.607                  & 0.361               & 0.597 & 0.656            \\
 & & (0.626)        & (0.533)  & (0.625) & (0.629)         \\
\textbf{\textit{\texttt{GPT4}}}                               & \textit{\textbf{8 x 220B}}                     & \textit{\textbf{0.761}}                 & \textit{\textbf{0.590}}           & \textit{\textbf{0.745}} &\textit{\textbf{0.781}}            \\ 
 & & (0.674)        & (0.583)  & (0.674) & (0.674)         \\
 \hline
\multicolumn{7}{|c|}{\textbf{Gradient-based Learning}}                                                                       \\ \hline
\texttt{RoBERTa}                                  & 380M                      & 0.464                  & 0.000               & 0.433 & 0.500             \\
 & & \textit{\textbf{(0.7916)}}        & \textit{\textbf{(0.75)}}  & \textit{\textbf{(0.787)}} & \textit{\textbf{(0.813)}}         \\
\texttt{MentalRoBERTa}          & 110M & 0.566                   & 0.200                      & 0.773                  & 0.554           \\ 
 & & (0.6)        & (0.4)  & (0.686) & (0.604)         \\
 \hline      
\multicolumn{7}{|c|}{\textbf{Anomaly Detection}}  \\   \hline  
\texttt{ECOD}                                  & N/A                      & 0.490                 & 0.08              & 0.496 & 0.498            \\
 & & (0.64285)        & (0.686)  & (0.5) & (0.666)          \\
\end{tabular}
\caption{Performance of models in detecting video-based MHMisinfo content. Bold text represents the best performance for the metric among open-source LLMs (that is, excluding the GPT models). Bold and italic text represent the best performance across all models. Bitchute-related metrics are presented with round brackets} \label{tbl:model_perf}
\end{table}

We compare the effectiveness of 5-shot ICL-trained models to zero-shot trained models where a prompting strategy to generate the prediction is also used. Unlike 5-shot learning, no demonstrations are given to the model in the zero-shot setting. We also use \texttt{RoBERTa} and \texttt{MentalRoBERTa} as gradient learning baselines \cite{liu2019roberta, ji2022mentalbert}. We chose these models due to their usage in previous digital mental health studies \cite{yang2023towards, zhang2022symptom}. Finally, given the class imbalance between MHMisinfo and non-MHMisinfo videos, we also tested the effectiveness of anomaly detection baselines in detecting MHMisinfo videos with the state-of-the-art \texttt{ECOD }algorithm, which has been shown to significantly outperform traditional anomaly detection methods such as one-class SVM in a variety of tasks \cite{li2022ecod}. For both gradient learning and anomaly detection baseline models, we train them using a fixed 80-20 train-test split under a stratified regime such that the ratio of non-MHMisinfo to MHMisinfo videos is similar between the splits for both Bitchute and YouTube data. All experiments are carried out on a compute cluster with 5 x Nvidia A100

\subsection{Results}
We report the results of all classification experiments for both the labeled YouTube and Bitchute datasets in Table \ref{tbl:model_perf}, where Bitchute results are presented within round brackets and YouTube results without.  We found that across the board, few-shot LLMs outperformed their zero-shot learning counterparts by a substantial margin. This ranges from a 1.47\% increase in macro-F1 for the \texttt{Flan-T5} model to a 60.5\% increase for the \texttt{Mistral-0.1} model. For the labeled YouTube videos, we found that most few-shot ICL-trained LLMs also out-performed the best gradient-based baseline (\texttt{MentalRoBERTa}), with the best model (\texttt{GPT4}) outperforming it by 0.2 F1-Macro and 0.39 F1-Positive. In contrast, for the labeled Bitchute videos, we found that even the best few-shot ICL-trained LLMs do not outperform the best gradient learning baseline (\texttt{RoBERTa}), with the best model (\texttt{GPT4}) underperforming it by 0.12 F1-Macro and 0.17 F1-Positive. While the best-performing LLM, that being \texttt{GPT4}, is also the largest among those tested, it is not always the case that larger models perform better than smaller models in detecting mental health misinformation (MHMisinfo). For instance, \texttt{Flan-T5} with only around 3 billion parameters outperformed the 175-billion-parameter \texttt{GPT3.5} in detecting YouTube MHMisinfo videos while being both substantially faster and less compute-insensitive. We use the best model for each dataset, \texttt{GPT4} for YouTube and RoBERTa for Bitchute to label all unlabeled YouTube ($n = 6658$) and Bitchute ($n = 887$) videos with at least 1 comment. We will now refer to the dataset containing both human-labeled videos from the \texttt{MentalMisinfo} dataset and the machine-labeled videos here as \texttt{MentalMisinfo-Large}. All results presented in the forthcoming sections are generated using the human-labeled \texttt{MentalMisinfo} dataset unless stated otherwise. \footnote{Both \texttt{MentalMisinfo} and \texttt{MentalMisinfo-Large} datasets, alongside all code needed to reproduce results from this paper can be found at \underline{\url{https://tinyurl.com/yrt7kaut}}}

\section{RQ 1: Linguistic Analysis of Engagement with Mental Health Misinformation}

\subsection{Engagement Analysis}
To understand how engagement to MHMisinfo and non-MHMisinfo videos differ on YouTube and Bitchute, we first analyze relevant video-level metadata that we have previously retrieved. These include the number of comments, the number of likes, the number of views, and the like-to-view ratio.  We use a statistical $t$-test for comparisons, utilizing Welch’s $t$-test to account for the imbalance of MHMisinfo and non-MHMisinfo videos. For YouTube, we find that while non-MHMisinfo videos receive significantly more views ($p = 0.035$) and comments ($p = 0.012$) than MHMisinfo videos, MHMisinfo videos have a significantly higher like-to-view ratio ($p = 0.047$) while also having no significant difference in like count ($p > 0.05$). We observed that non-MHMisinfo Bitchute videos also received more comments compared to MHMisinfo Bitchute videos ($p < 0.005$)

\begin{table}[]
    \centering
    \small
    \setlength{\tabcolsep}{0.5pt}
    \begin{tabular}{l@{}|ll|l|l|l}
        \textbf{LIWC}    & \textbf{NMisinfo(\%)} & \textbf{Misinfo(\%)} & \textbf{\%Diff} & \textbf{$q$-val}             & \textbf{$d$} \\ \hline
        \multicolumn{6}{|c|}{\textbf{Affect}}                                                                                                                                              \\ \hline
       \texttt{affect} %(affect)            
        & 6.3                         & 6.13                            & -2.69                  &                         & 0.02               \\
         
        & (6.8)                          & (6.0)                             & (-11.48)                  &                         & (0.07)               \\
        \texttt{posemo} %(posemo)  
        & 3.79                        & 3.64                            & -4.06                  &                         & 0.02               \\
        & (3.0)                          & (2.9)                             & (-4.46)                  &                         & (0.016)               \\
        \texttt{negemo} %(negemo)  
        & 2.45                        & 2.45                            & 0.00                   &                        & 0.00               \\
        & (3.7)                          & (2.99)                             & (-19.44)                  &                         & (0.093)               \\
        \texttt{anxiety} %(anx)              
        & 0.53                       & 0.26                            & -50.01                  & *** (§)   & 0.11                \\
        & (0.28)                        & (0.08)                           & (-69.91)                 & ** (§)                        & (0.128)          \\
        \texttt{sadness} %(sad)              
        & 0.56                       & 0.94                         & 68.29                & *** (†) & -0.14             \\ 
        & (0.43)                          & (0.67)                             & (56.07)                  &                         & (-0.062)               \\ \hline
        \multicolumn{6}{|c|}{\textbf{Social}}                                                                                                                                              \\ \hline
        \texttt{social} %(Social)           
        & 8.39                         & 9.15                          & 9.16                   & *** (†)  & -0.07               \\
        & (7.6)                          & (6.9)                             & (-9.51)                  &                         & (0.08)               \\
        \texttt{family} %(Family)            
        & 0.82                        & 0.60                           & -27.24               & *** (§)   & 0.06               \\
        & (0.28)                          & (0.34)                             & (17.42)                  &                         & (-0.028)               \\
        \texttt{friend} %(Friends)           
        & 0.26                         & 0.32                           & 21.43                   & * (†)   & -0.03               \\
        & (0.24)                          & (0.22)                             & (-8.541)                  &                        & (0.013)               \\
        \texttt{female} %(Female)            
        & 1.28                         & 0.29                            & -77.78                   & *** (§)                           & 0.23                 \\
        & (0.9)                             & (0.3)                                & (-60.58)                    & ** (§)                          & (0.172)                \\
        \texttt{male} %(Male)                
        & 0.87                         & 2.06                             & 137.13                 & *** (†) & -0.32               \\ 
        & (1.06)                            & (0.67)                               & (-37.29)                  & * (§)                         & (0.113)                \\ \hline
        \multicolumn{6}{|c|}{\textbf{Somatic}}                                                                                                                                             \\ \hline
        \texttt{bio} %(Biological Processes) 
        & 3.06                        & 2.61                            & -14.61                  & *** (§) & 0.07                \\
        & (2.56)                          & (2.39)                             & (-6.92)                  &                         & (0.028)               \\
        \texttt{body} %(Body)                
        & 0.62                        & 0.69                             & 11.36                  &                        & -0.03              \\
        & (0.9)                          & (0.6)                             & (-34.35)                  &                         & (0.1)               \\
        \texttt{health} %(Health)            
        & 1.68                         & 1.38                            & -18.11                   & *** (§)  & 0.06               \\
        & (0.86)                         & (1.36)                            & (58.23)                  & * (†)                        & (-0.12)               \\
        \texttt{sexual} %(Sexual)           
        & 0.11                        & 0.17                           & 53.05                 & ** (†) & -0.04              \\
        & (0.56)                         & (0.17)                            & (-69.93)                  & * (§)                        & (0.02)               \\
        \texttt{ingest} %(Ingest)            
        & 0.45                         & 0.37                           & -18.20                 & *** (§) & 0.03               \\ 
        & (0.33)                          & (0.27)                             & (-17.58)                  &                         & (0.034)               \\ \hline
        \multicolumn{6}{|c|}{\textbf{Drives}}                                                                                                                                              \\ \hline
        \texttt{drives} %(Drives)           
        & 4.50                       & 6.33                            & 40.79                   & *** (†) & -0.25              \\
        & (4.9)                          & (4.7)                             & (-2.69)                  &                         & (0.02)               \\
        \texttt{affl.} %(Affiliation)  
        & 1.39                         & 1.52                           & 9.00                   & * (†)                         & -0.03               \\
        & (0.9)                          & (1.0)                             & (8.38)                  &                         & (-0.026)               \\
        \texttt{achieve} %(Achievement)     
        & 0.68                       & 1.17                            & 71.00                   & *** (†) & -0.18               \\
        & (0.86)                          & (0.85)                             & (-2.82)                  &                       & (0.02)               \\
        \texttt{power} %(Power)              
        & 1.33                         & 2.41                             & 82.09                  & *** (†) & -0.27              \\
        & (2.1)                          & (1.6)                             & (-22.23)                  &                         & (0.105)               \\
        \texttt{reward} %(Reward)            
        & 1.06                        & 1.12                            & 4.91                   &                         & -0.02              \\
        & (1.11)                          & (1.10)                             & (-4.33)                  &                         & (0.0133)               \\
        \texttt{risk} %(Risk)                
        & 0.44                        & 0.76                             & 72.66                  & *** (†) & -0.15               \\ 
        & (0.39)                          & (0.56)                             & (40.01)                  &                        & (-0.098)               \\ \hline
        \multicolumn{6}{|c|}{\textbf{Personal Concerns}}                                                                                                                                  \\ \hline
        \texttt{work} %(Work)                
        & 0.69                         & 0.99                             & 43.08                   & ***(†) & -0.11               \\
        & (1.6)                          & (1.78)                             & (11.81)                  &                         & (-0.048)               \\
        \texttt{leisure} %(Leisure)          
        & 0.72                        & 1.24                            & 72.09                   & *** (†) & -0.16               \\
        & (0.9)                          & (1.2)                             & (40.89)                  &                         & (-0.123)               \\
        \texttt{home} %(Home)                
        & 0.22                         & 0.12                            & -46.38                 & *** (§)& 0.07             \\
        & (0.27)                          & (0.26)                             & (-2.39)                  &                         & (0.0037)               \\
        \texttt{money} %(Money)              
        & 0.16                         & 0.23                           & 46.22                   & *** (†)                        & -0.05             \\
        & (0.33)                          & (0.19)                             & (-41.63)                  &                         & (0.097)               \\
        \texttt{religion} %(Religion)           
        & 0.38                        & 1.52                            & 300.98                 & ***  & -0.31             \\
        & (0.4)                          & (0.43)                             & (8.29)                  &                        & (-0.015)               \\
        \texttt{death} %(Death)              
        & 0.10                         & 0.16                            & 57.20                  & *** (†) & -0.06              \\ 
        & (0.28)                          & (0.48)                             & (72.61)                  &                        & (-0.126)               \\ \hline         
    \end{tabular}
    \caption{Relevant LIWC categories in the comments, along with results of Student's $t$-tests. Rows with section marks (§) following the q-value indicate that usage among non-MHMisinfo (NMisinfo) comments is significantly higher. Rows with daggers (†) following q-value indicate that usage among MHMisinfo (Misinfo) comments is significantly higher. Bitchute-related metrics are presented with round brackets. * indicates \(q < 0.05\), ** indicates \(q < 0.01\), *** indicates \(q < 0.001\)}
    \label{tbl:liwc}
\end{table}

\subsection{Linguistic Analysis}
\subsubsection{Methods}
To understand who engages with MHMisinfo videos on mental health and why, we use the Linguistic Inquiry and Word Count (LIWC) analysis program \cite{boyd2022development}. LIWC is well-validated for analyzing social media data, especially in mental health contexts \cite{de2014mental, pendse2023marginalization}. Our analysis focuses on LIWC categories related to perceptions about mental health conditions, including language about affect, community, the body, religion, class, and gender. We remove all stopwords using the Natural Language Toolkit stopword list and normalize all counts by the length of the comment. We use Welch's $t$-test to measure significant differences in lexicon usage across these relevant categories per platform, with False Discovery Rate (FDR) adjustment at a significance level of 0.05, denoted as the $q$-value. In Table \ref{tbl:liwc}, we present the average percentage of a comment containing language for each LIWC category for both MHMisinfo and non-MHMisinfo comments on both platforms, the percentage difference in usage within MHMisinfo comments compared to non-MHMisinfo comments, and the corresponding $q$-values and Cohen's $d$.

\begin{table}[]
\centering
\small 
\setlength{\tabcolsep}{1.7pt}
\begin{tabular}{c|lrlr}
\textbf{Platform} & \multicolumn{2}{l|}{\textbf{Misinfo}}           & \multicolumn{2}{l|}{\textbf{NMisinfo}} \\ \hline 
& $n$-gram & SAGE & $n$-gram        & SAGE       \\ \hline
& schizophrenia               & 2.373                    & alice           & -3.951                                                           \\
& jesus                       & 2.323                    & relatable           & -3.768                              \\
& pray                        & 2.196                    & birthday           & -3.767                              \\
& step                        & 2.132                    & stim           & -3.744                              \\
& lord                        & 1.943                    & adorable           & -3.495                              \\
& christ                      & 1.828                    & julian          & -3.292                              \\
& depression                  & 1.653                    & mama           & -3.272                              \\
& holy                        & 1.641                    & princess           & -3.249                              \\
& amen                        & 1.614                    & relate           & -3.192                              \\
& free                        & 1.447                    & sweet           & -3.167                              \\
& medication                  & 1.423                   & ariel           & -3.144                            \\
\textbf{YouTube} & depressed                   & 1.422                    & functioning           & -2.678                            \\
& meds                        & 1.402                    & adhd           & -2.506                            \\
& truth                       & 1.313                    & baby           & -2.418                              \\
& cure                        & 1.283                    & hair           & -2.339                              \\
& name                        & 1.195                    & mens           & -2.308                              \\
& working                     & 1.191                    & wall          & -2.239                              \\
& believe                     & 1.088                    & precious          & -2.144                             \\
& word                        & 1.065                    & accurate           & -2.051                              \\
& real                        & 0.993                    & cute           & -2.048                              \\
& father                      & 0.985                    & awareness           & -1.854                              \\
& drugs                       & 0.961                    & asian          & -1.853                              \\
& trans                       & 0.959                    & cleaning           & -1.819                              \\
& agree                       & 0.952                    & teacher           & -1.779                             \\
& stop                        & 0.943                    & tired           & -1.764                               \\ \hline                       
& adhd              & 0.953                    & mental           & 0.375                                                           \\
& nicotine                       & 0.869                    & ni***rs           & 0.260                              \\
& aluminum                        & 0.869                    & looks           & 0.246                              \\
& generations                        & 0.847                    & health           & 0.240                              \\
& sources                       & 0.816                    & whites           & 0.225                              \\
& pfizer                      & 0.816                    & women          & 0.203                              \\
& doctor                  & 0.8                    & jews           & 0.189                              \\
& chief                        & 0.775                    & right           & 0.007                              \\
& treatment                        & 0.775                    & like           & 0.0002                              \\
& debates                        & 0.769                    & thats           & 0.0001                              \\
& nuremburg                 & 0.744                   & white           & 0.0001                            \\
\textbf{Bitchute} & doctors                   & 0.744                    & need           & 0.0001                            \\
& news                        & 0.707                    & even           & 0.0001                            \\
& 1986                       & 0.703                    & look           & \(9.22 * 10^{-5}\)                              \\
& parents                        & 0.673                    & stop           & \(9.01 * 10^{-5}\)                              \\
& religious                        & 0.642                    & someone           & \(8.81 * 10^{-5}\)                          \\
& medical                     & 0.642                    & well          & \(8.49 * 10^{-5}\)                              \\
& effective                     & 0.642                    & race          & \(8.19 * 10^{-5}\)                             \\
& cult                        & 0.627                    & b***h           & \(7.66 * 10^{-5}\)                              \\
& autism                        & 0.585                    & thing           & \(7.54 * 10^{-5}\)                              \\
& true                      & 0.579                    & real           & \(7.54 * 10^{-5}\)                              \\
& generation                       & 0.563                    & guilt          & \(7.46 * 10^{-5}\)                              \\
& diagnosed                       & 0.563                    & blacks           & \(7.46 * 10^{-5}\)                              \\
& believe                       & 0.549                    & shes           & \(7.46 * 10^{-5}\)                             \\
& childhood                        & 0.534                    & yeah           & \(7.32 * 10^{-5}\)                              \\ 
\hline
\end{tabular}
\caption{Top-25 most discriminative n-grams for MHMisinfo (Misinfo) comments and non-MHMisinfo (NMisinfo) comments for YouTube and Bitchute, identified through the SAGE algorithm} \label{tbl:sage}
\end{table}
In addition to LIWC, we use an unsupervised language modeling technique called the Sparse Additive Generative Model (SAGE) introduced by Eisenstein et al. (2011) \cite{eisenstein2011sparse}. We employ SAGE to identify discriminating $n$-grams ($n = 1, 2, 3$) between comments on MHMisinfo and non-MHMisinfo videos. The magnitude of the SAGE value of a linguistic token indicates the degree of its uniqueness to the distribution of comments on MHMisinfo videos relative to the distribution of comments on non-MHMisinfo videos. For each word, a positive SAGE value greater than 0 suggests that the $n$-gram is more representative of comments on MHMisinfo videos, while a negative SAGE value suggests greater representativeness for comments on non-MHMisinfo videos. We initialize the SAGE model using the 500 most frequent words (excluding stopwords and words shorter than 3 characters) for both MHMisinfo and non-MHMisinfo comments, with a baseline smoothing of 1. Table \ref{tbl:sage} presents the top $n$-grams for both MHMisinfo and non-MHMisinfo comments on both platforms.

\subsubsection{Results}
\paragraph{YouTube}
As we present in Table \ref{tbl:liwc}, we find that commenters to MHMisinfo videos are less likely to use some types of somatic language compared to commenters to non-MHMisinfo comments, with language related to health ($q$ \(< 0.001\)) being used more in response to non-MHMisinfo videos. This may imply that commenters to non-MHMisinfo videos are more likely to be aware of and concerned about the effect of mental disorders on one's overall health:

\begin{quote}
    \textit{It’s \textbf{physically}
and emotionally \textbf{painful}. I feel so bad} (bolded words are in LIWC's \texttt{health} category)
\end{quote}

%\newline\newline
In contrast, language related to \texttt{risk} ($q$ \(< 0.001\)) and \texttt{death} ($q$ \(< 0.001\)) are significantly more used in MHMisinfo comments. This is likely connected to the tendency professed by many of these comments to frame the impact of mental health conditions with one's mortality:

\begin{quote}
    \textit{'Yeah, they had depression, but if they showed it they were useless to others, so they were left to \textbf{die}.} (the bolded word is in LIWC's \texttt{body} category)
\end{quote}

Self-perceptions about one's gender and ambitions also play a key role in how people react to MHMisinfo videos. We find significantly higher usages of \texttt{male} words within MHMisinfo comments ($q$ \(< 0.001\)), whereas non-MHMisinfo comments have significantly higher usages of female-referent words. Commenters to MHMisinfo videos also utilize more words related to \texttt{power} and \texttt{achievement} such as ``defeat" or ``strength". This seems to suggest the people consuming and responding to MHMisinfo videos are more likely to be male and more likely to believe in the analogy that mental disorders are ``enemies" needing to be defeated and that people who suffer from mental disorders are not strong enough to defeat the ``enemy":

\begin{quote}
    \textit{Imagine have depression... as wise Andrew Tate said, just a \textbf{weak} mindset.} (bolded words are in LIWC's \texttt{power} category)
\end{quote}

Finally, we also find that comments to MHMisinfo videos are 4 times more likely to evoke \texttt{religion} language when compared to comments to non-MHMisinfo videos. ($q$ \(< 0.001\)). This is further corroborated by the most distinguishing words deployed by the former identified by SAGE, with 6 of the top ten SAGE keywords being religion-related: ``jesus", ``pray", ``lord", ``christ", ``holy", and ``amen". Both evidence point to a substantial portion of users who engage with mental health misinformation having a faith-centered understanding of mental ailments and treatments and are more likely to distrust evidence-based therapies:
% \begin{quote}
%     \textit{After a \textbf{satanic cult} \textbf{ritualized} me I began having schizophrenic symptoms. \textbf{Jesus} delivered from them all and I am healed today. } (bolded words are in LIWC's religion category)
% \end{quote}
\begin{quote}
    \textit{Got put on medication that did nothing…. Got into the Word, \textbf{prayer}, \textbf{worship} and acts of service—it all fled. Hallelujah}
\end{quote}
\begin{quote}
    \textit{It's hard, but turning to \textbf{Christ} will always be the way out} (bolded words are in LIWC's \texttt{religion} category)
\end{quote}
We found that MHMisinfo and non-MHMisinfo comments follow similar LIWC and SAGE score patterns even when accounting for machine-labeled videos and comments in the MentalMisinfo-Large dataset. Such results are shown in tables \ref{tbl:liwc_appendix} and \ref{tbl:sage_appendix} of the Appendix section.  
\paragraph{Bitchute}
However, there are few significant differences between MHMisinfo and non-MHMisinfo Bitchute comments. This could be partly attributed to Bitchute's alt-tech nature, which attracts users with more homogeneous viewpoints regarding mental health compared to YouTube. Although some LIWC features differ between these sets of comments, some differences are opposite in direction from their corresponding sets on YouTube. For instance, comments responding to MHMisinfo videos on Bitchute show significantly higher usage of language related to health and lower usage of language related to masculinity and sexuality. An inspection of labeled Bitchute videos reveals that much of these differences are driven by the prevalence of anti-transgender content within videos labeled as non-MHMisinfo. Despite substantial differences in user demographics and mental health content posted on these platforms, there are still similarities in linguistic patterns of comments on MHMisinfo videos. Notably, treatment-related (e.g., \textit{medication} for YouTube, \textit{treatment} for Bitchute) and religion-related (e.g., \textit{christ} for YouTube, \textit{religious} for Bitchute) SAGE n-grams persist across both platforms.

\section{RQ 2: Assessing Agreement to Mental Health Misinformation}

We model the agreement detection task as follows: Given a video's full transcript (which is created by combining the video's title, and audio transcript with its on-screen text content i.e. visual transcript and special delimiting characters surrounding each of the transcripts), and the full text of the comment, we intend to determine whether the comment ``agrees" or ``does not agree" with the video. Here, ``does not agree" can either indicate disagreement, neutrality, or that the comment is irrelevant to the video (i.e. spam). 

To measure this, similar to RQ2, we built a Large Language Model (LLM)-based agreement classifier using a few-shot in-context learning approach. We randomly sample comments from the labeled dataset and label them for agreement, until we get a sample of 3 comments agreeing to the video and 2 comments not agreeing to the video. Given the non-specialized nature of the task unlike other tasks in this paper which involve specialized knowledge regarding mental health, we use the \texttt{Mistral} model as our base LLM. We feed \texttt{Mistral} 5 examples and their agreement label, followed by the target comment and its associated video. Then we ask it to generate the agreement label for such video-comment pair. To verify the validity of the classifier for large-scale labeling, the first author and the supervising co-author randomly labeled 100 video-comment pairs for ground truth agreement. We compare the machine-generated agreement labels with the ground truth agreement, which yields a high F1 score of 0.86. % The experiment was performed using a single Nvidia A100 GPU, and it took us 184 GPU hours to label all 134420 comments for agreement.

%\subsubsection{Results}
We statistically compare the distribution of ``agree" to ``does not agree" comments within responses to MHMisinfo and non-MHMisinfo videos using the $\chi^2$ test. %, with the null hypothesis being that they are similarly distributed. 
For YouTube, we found that comments in response to MHMisinfo videos are significantly less likely to agree with their content (\(\%_{agreement}\) = 5.77\%) than those in response to non-MHMisinfo videos (\(\%_{agreement}\) = 20.14\%) ($p<0.001$). The trend holds even when we do the comparison on the YouTube videos in the \texttt{MentalMisinfo-Large} dataset ($p<0.001$). In contrast, we found the opposite trend for comments in response to Bitchute MHMisinfo videos, where there is significantly higher levels of video-comment concordance there (\(\%_{agreement}\) = 15.5\%) compared to non-MHMisinfo videos (\(\%_{agreement}\) = 6.2\%). These agreement trends hold even if we tested on the \texttt{MentalMisinfo-Large} dataset. We present results regarding video-comment agreement for both YouTube and Bitchute in Table \ref{tbl:rq23_comments} for the \texttt{MentalMisinfo} dataset and Table \ref{tbl:rq23_comments_appendix} of the Appendix section for the \texttt{MentalMisinfo-Large} dataset.

\begin{table}[t]
\small
\begin{tabular}{l@{}l@{}l@{}l@{}l@{}l}
Platform & \,\,Metric\,\,  & $\%$ Misinfo & \,\,$\%$  NMisinfo\,\, & p-value\,\, & \(\chi^2\) \\ \hline
YouTube & \,\,Agreement\,\, & 5.8$\%$  & \,\,20.1$\%$ & ***          & $4.8 * 10^4$                      \\
& \,\,Stigma\,\,    & 32.9$\%$         & \,\,21.9$\%$           &   ***           & 6948.37                      \\ \hline     
Bitchute & \,\,Agreement\,\, & 15.5$\%$  & \,\,6.2$\%$  & ***          & 572.32                     \\
& \,\,Stigma\,\,     &   53.4$\%$    & \,\, 49.3$\%$         &    *           & 5.2                      \\ \hline   
\end{tabular}
\caption{Results summary for RQs 2 (Agreement Analysis) and 3 (Stigma Analysis) on YouTube and Bitchute comments. Here, \textsuperscript{*} indicates \(p < 0.05\), \textsuperscript{**} indicates \(p < 0.01\), \textsuperscript{***} indicates \(p < 0.001\)}\label{tbl:rq23_comments}
\end{table}

\section{RQ 3: Assessing Stigma in Responses to Mental Health Misinformation}
Finally, we are interested in whether comments in response to mental health misinformation (MHMisinfo) videos are more stigmatizing when compared to responses to non-MHMisinfo videos. Previous evidence from literature has pointed to a dangerous potential feedback loop between mental health stigma and misinformation related to various mental health disorders \cite{corrigan2004stigma}. If left uncontrolled, this feedback loop can cause serious harm to people's mental well-being and degrade trust in the medical system. Therefore, it is important to verify whether evidence of heightened levels of stigma linked to mental health misinformation can be found on consequential video-sharing platforms. % like YouTube.

%\subsection{Methods}
We frame the stigmatizing comment detection tasks as a binary detection task. Given a comment $c = \{w_1, w_2, ..., w_n\}$ containing $w_i$ words, we trained a classifier $F$ that learns the function $F(c) \longrightarrow y \in \{0, 1\}$, where $y$ is the stigma label of the comment (0 for no stigma, 1 for stigma). Similar to previous classification approaches presented in this paper, we utilize 5-shot ICL to train this classifier $F$. To generate the data used for few-shot ICL training, we randomly sample comments from \texttt{MentalMisinfo} fo  Due to the specific and complex nature of mental health stigma and its linguistic manifestations, we use the specialized \texttt{MentaLLaMa-13B} model as our base LLM \cite{yang2023towards}. We used the prompt template given in Table \ref{tbl:prompt_stigma_appendix} as the input.

Similar to the agreement classifier, we verify the performance of the classifier by randomly labeling 100 comments manually for ground truth, drawing on existing stigma literatures~\cite{mittal2023moral}. % We then compare the machine-generated labels for these 100 comments against the ground truth, which 
Our approach yields a satisfactory F1 score of 0.875. % It took us 279 GPU hours to label all 134420 comments for stigma on a single V100 GPU
%\subsubsection{Results}
Similar to the agreement analysis, we statistically compare the distribution of stigmatizing to non-stigmatizing comments within responses to MHMisinfo and non-MHMisinfo videos using the $\chi^2$ test, %with the null hypothesis being that they are similarly distributed. 
We found for both Bitchute and YouTube, comments in response to MHMisinfo videos contain significantly more stigmatizing views against mental health compared to those of non-MHMisinfo videos. The gap in stigma presence within MHMisinfo comments and non-MHMisinfo comments is greater between YouTube comments (32.92\% vs. 21.9\%) compared to Bitchute comments (53.4\% vs. 49.35\%), most likely due to Bitchute commenters being more likely to hold stigmatizing views regarding mental health whether they are exposed to MHMisinfo content or not. As is the case with our agreement analysis in RQ2, the trends hold even when we repeated the experiment on the \texttt{MentalMisinfo-Large} dataset. We present results regarding stigma-holding comments for both YouTube and Bitchute in Table \ref{tbl:rq23_comments} for the \texttt{MentalMisinfo} dataset and Table \ref{tbl:rq23_comments_appendix} of the Appendix section for the \texttt{MentalMisinfo-Large} dataset.

\section{Discussion}
\subsection{Implications for Misinformation Research}

Our research provides an empirical understanding of user engagement with short-form mental health misinformation (MHMisinfo) on video-sharing platforms, focusing on automated detection techniques. We introduce the \textit{first dataset} and \textit{framework} for classifying online mental health (MH) misinformation. Leveraging the task generalization capabilities of LLMs, we demonstrate their superiority over gradient-based learning approaches, especially in few-shot in-context learning. Consistent with recent work by Ziems et al. (2023), our study shows that models like \texttt{GPT-4} and \texttt{Mistral-0.1} significantly outperform baseline models in detecting MHMisinfo on YouTube. However, while few-shot approaches underperform compared to baselines in detecting Bitchute MHMisinfo, they still offer robust identification of a wide variety of MHMisinfo across different platforms with very few labeled examples, allowing for natural language explanations of the classifier's decisions. This suggests new possibilities for addressing online misinformation challenges with advances in generative AI.

Unlike prior research, our study goes beyond demonstrating that AI techniques can detect online misinformation, exploring how this misinformed content is perceived in the broader social media ecosystem. Analysis of our unique MHMisinfo dataset reveals significant linguistic disparities in reactions to MHMisinfo versus non-MHMisinfo content. On YouTube, our quantitative analysis of comments indicates that audiences responding to MHMisinfo videos are more likely to be male, religious, and inclined to distrust evidence-based approaches to mental health care. This aligns with previous research on susceptibility to online health misinformation (Nan et al., 2022). Using an LLM-based approach, we observe lower agreement rates with MHMisinfo content compared to non-MHMisinfo content, although the presence of any agreement is alarming. Even more concerning are our observations on Bitchute comments, where there is substantially higher agreement with MHMisinfo videos compared to non-MHMisinfo videos. Such rates of agreement likely reflect the site's population, which is more prone to conspiratorial thinking. Additionally, MHMisinfo comments on both YouTube and Bitchute show a higher likelihood of perpetuating stigma against individuals with mental health conditions. Our findings highlight the urgent need to not only tackle MHMisinfo content through algorithmic strategies but also to implement approaches to reduce the resulting harm.

We make recommendations for how online platforms can design better pathways to high-quality mental health information. Furthermore, we discuss how public health personnel can effectively utilize online social media platforms to address common mental health misconceptions and disseminate adaptive mental health resources to better meet the needs of those who may lack access to care or be marginalized.
\subsection{Platform Design Recommendations}
\subsubsection{Adaptive Interventions against Mental Health Misinformation}
Our work quantitatively demonstrates how certain mental health-related topics are susceptible to MHMisinfo content. Additionally, we show that certain individual characteristics, such as male and religious-inclined users, are more likely to be receptive to misinformation within these topics.

To combat the spread of MHMisinfo, platforms can restrict access to such content by demoting it from users' search recommendations or removing it altogether \cite{van2022misinformation}. Large language models (LLMs) offer a powerful solution for content moderation efforts due to their general capabilities and adaptability to domain-specific tasks. LLMs can identify MHMisinfo without the need for labor-intensive dataset curation required by gradient-based modeling approaches. However, it's important to acknowledge that these models aren't perfect, and there's a risk of misclassification. To mitigate this risk, platforms can incorporate human moderators into the framework and resort to content removal only in clear-cut cases. They can also implement additional methods to identify suspected MHMisinfo content, such as harnessing community reporting, as utilized by Twitter's erstwhile Community Notes program \cite{prollochs2022community}.

Despite efforts to moderate MHMisinfo, it remains challenging to monitor every piece of content, and susceptible users can still seek out, consume, and share MHMisinfo. Therefore, platform-based interventions should focus on empowering susceptible users to combat MHMisinfo and access relevant mental health resources. Promising mechanisms include credibility markers for health professionals (prominently deployed by YouTube) and keyword-triggered verified resource labels for specific mental health conditions (prominently deployed by TikTok). However, these interventions often lack demographic and geographic context, missing the opportunity to provide health information resources more likely to be adopted by susceptible users. For example, users who self-identify as religious could be provided with faith-based mental health resources advocating for the integration of evidence-based and spiritual practices in mental health care. Platforms could also adopt a Community Notes-style system for individual mental health-related videos, inviting members to fact-check or provide additional context.

\subsubsection{Pathways from Online Information to Offline Care}
The counterspeech doctrine, as established by Justice Brandeis in the 1927 landmark case Whitney v. California, expressed the idea that the preferred solution to bad speech in the framework of the First Amendment should be more speech \cite{richards2000counterspeech}. While the effectiveness of doctrine has been questioned in recent years due to the rapid spread of misinformation and hate speech across social media platforms, high-quality credible information on mental health is still needed to meet the needs of viewers on all stages of their mental health journey \cite{gowen2013online}. Mental health professionals such as psychiatrists or licensed therapists are the most well-positioned to make accurate and evidence-based claims on mental health conditions \cite{kutcher2016mental}. However, given their other commitments and the widely acknowledged paucity of trained professionals relative to public health needs~\cite{butryn2017shortage}, platforms should aim to lower the barrier to creating high-quality yet accessible videos for verified mental health experts. Practical solutions for achieving this goal might include providing free video templates, editorial assistance, and algorithmically promoting content produced by mental health professionals. 

Our findings also echo previous research on how most popular mental health-related content on social media platforms promote non-nuanced, non-agentive narratives on mental illness over more nuanced and agentive narratives \cite{milton2023see}. The former are much more likely to be MHMisinfo based on our criteria, as it discourages the usage of evidence-based treatment and has low alignment with medical consensus on mental health. Platforms could encourage the latter through algorithmic promotion and platform-creator partnerships in a similar vein to the YouTube Partner Program\footnote{\underline{\url{https://support.google.com/youtube/answer/72851}}} to incentivize the creation of high-quality patient-centered mental health content. 

\subsection{Limitations and Future Work}
Our quantitative research on mental health misinformation (MHMisinfo), the first of its kind, identifies constraints that pave the way for future development in this critical area. Focused on monolingual content and responses (English) from YouTube and Bitchute, our study urges further exploration of cross-cultural differences in mental health narratives \cite{%de2017gender, 
pendse2019cross}. % Investigating how culture shapes the prevalence of misinformative mental health content, engagement, and interventions is essential. 
A vital avenue also includes understanding MHMisinfo on youth-centric platforms like TikTok and Instagram, given the heightened susceptibility of adolescents \cite{patel2007mental}. Future research could pursue such avenue given adequate regulatory permission. 
Further, our linguistic findings revealed that users engaging with MHMisinfo content are often male and faith-oriented. Due to limitations in inferring demographic information through existing data collection mechanisms, future research could directly recruit participants who utilize video-sharing platforms for their mental health informational needs and conduct population-level analyses with user-contributed engagement data. Controlled experiments with these participants could illuminate attitudes towards MHMisinfo and identify effective interventions for MHMisinfo.

\section{Conclusion}
Mental health resources diffused through online video-sharing platforms have the potential to substantially augment traditional mental health services, especially in MPSAs in the United States and around the world. However, mental health misinformation can substantially exacerbate stigma towards individuals afflicted with mental health and harm afflicted individuals through ignorance, or even rejection of effective interventions. Therefore, mental health professionals, researchers, and even online platforms must understand mental health misinformation (MHMisinfo) content and community engagement to such content. %, so effective interventions against such content can be devised. 
With our novel \texttt{MentalMisinfo} dataset, our paper first demonstrates the effectiveness of large language models (LLMs) under few-shot in-context learning in detecting MHMisinfo content. More central to our research focus, we deployed novel LLM-based approaches in conjunction with well-validated linguistic analysis techniques to find many notable trends regarding engagement with MHMisinfo content. More specifically, we found that audiences responding to MHMisinfo videos are more likely to be male, religious, and inclined to distrust evidence-based approaches to mental health care. We also found that MHMisinfo videos generated substantial agreement from commenters, and in the case of Bitchute, significantly more when compared with agreement to non-MHMisinfo videos. Finally, comments in MHMisinfo videos were more highly associated with stigmatizing views. Our findings could inform potential adaptive interventions to counteract misinformation in favor of accurate and timely mental health information. We believe that our first large-scale quantitative analysis of MHMisinfo content on video-sharing platforms will enable future research on assessing and augmenting the quality of online mental health information.

\fontsize{9pt}{10pt} \selectfont
\bibliography{aaai24}

\section{Ethics Checklist}
\begin{enumerate}

\item For most authors...
\begin{enumerate}
    \item  Would answering this research question advance science without violating social contracts, such as violating privacy norms, perpetuating unfair profiling, exacerbating the socio-economic divide, or implying disrespect to societies or cultures?: \ethics{Yes. As discussed in the "Introduction" and "Discussion" sections, our paper advances the spaces of computational social science and digital mental health as the first to quantitatively examine mental health misinformation on video-sharing platforms. Our findings offer a balanced approach to detecting and moderating mental health misinformation with contextual consideration for freedom of speech rights and potential biases. Our nuanced approach aims to avoid profiling and violation of any relevant social contracts.}
  \item Do your main claims in the abstract and introduction accurately reflect the paper's contributions and scope?
    \ethics{Yes. We have carefully reviewed that the claims made in "Abstract" and "Introduction" sections have accurately reflect our contributions and project scope}
   \item Do you clarify how the proposed methodological approach is appropriate for the claims made? 
    \ethics{Yes (refer to the "Methods" subsection for RQ1, RQ2, RQ3 sections) We justify the appropriateness of our model and method selection in the \textit{Detection of Mental Health Misinformation} section by referring to prior works that use them for similar tasks/objectives. For RQ1, we justify our usage of specific LIWC categories and SAGE due to their usage in previous studies regarding online mental health discourse. In addition, we also quantitatively validate our LLM-based agreement and stigma classifiers in RQ2 and RQ3 by comparing a subset of machine-labeled comments to human-labeled contents}
   \item Do you clarify what are possible artifacts in the data used, given population-specific distributions?
    \ethics{Yes, we did specify potential artifacts in the YouTube and Bitchute data used in the "Datasets" section, such as those brought about by the set of keywords used.}
  \item Did you describe the limitations of your work?
    \ethics{Yes, we did discuss the limitations of our work in the "Limitations and Future Work" subsection within "Discussion"}
  \item Did you discuss any potential negative societal impacts of your work?
    \ethics{Yes, we did discuss the potential negative societal impacts of our work and how to mitigate them in the "Discussion" section)}
   \item Did you discuss any potential misuse of your work?
    \ethics{Yes, we did discuss the potential misuse of our work and how to mitigate them in the "Discussion" section)}
   \item Did you describe steps taken to prevent or mitigate potential negative outcomes of the research, such as data and model documentation, data anonymization, responsible release, access control, and the reproducibility of findings?
    \ethics{Yes, we did discuss the steps taken prevent to mitigate potential negative outcomes of the research, such as data anonymization, responsible release, and reproducibility of findings in the "Dataset" section.}
   \item Have you read the ethics review guidelines and ensured that your paper conforms to them?
    \ethics{Yes, we have read the ethics review guidelines and ensured that our paper conforms to them to the best of our capabilities}
\end{enumerate}

\item Additionally, if your study involves hypotheses testing...
\begin{enumerate}
  \item Did you clearly state the assumptions underlying all theoretical results?
    \red{NA}
  \item Have you provided justifications for all theoretical results?
    \red{NA}
  \item Did you discuss competing hypotheses or theories that might challenge or complement your theoretical results?
    \red{NA}
  \item Have you considered alternative mechanisms or explanations that might account for the same outcomes observed in your study?
    \red{NA}
  \item Did you address potential biases or limitations in your theoretical framework?
   \red{NA}
  \item Have you related your theoretical results to the existing literature in social science?
    \red{NA}
  \item Did you discuss the implications of your theoretical results for policy, practice, or further research in the social science domain?
    \red{NA}
\end{enumerate}

\item Additionally, if you are including theoretical proofs...
\begin{enumerate}
  \item Did you state the full set of assumptions of all theoretical results?
    \red{NA}
	\item Did you include complete proofs of all theoretical results?
    \red{NA}
\end{enumerate}

\item Additionally, if you ran machine learning experiments...
\begin{enumerate}
  \item Did you include the code, data, and instructions needed to reproduce the main experimental results (either in the supplemental material or as a URL)?
    \ethics{We have included the prompts used in LLM-based classifiers and experiments in the paper (refer to Methods subsection for RQ1, 2, and 3. We also plan to release the curated dataset publicly to enable replicability, pending the publication of this manuscript.} 
  \item Did you specify all the training details (e.g., data splits, hyperparameters, how they were chosen)?
    \ethics{Yes, we did include the prompts used for LLM-based experiments and training details for baselines in the Methods subsection for RQ1, 2, and 3}
     \item Did you report error bars (e.g., with respect to the random seed after running experiments multiple times)?
    \red{NA}
	\item Did you include the total amount of compute and the type of resources used (e.g., type of GPUs, internal cluster, or cloud provider)?
    \ethics{Yes, we include the total amount of compute and type of resources used for every LLM-based procedures (refer to the Methods subsection of sections \textit{Detection of Mental Health Misinformation}, \textit{RQ2: Analyzing Agreement with Mental Health Misinformation} and \textit{RQ3: Analyzing Stigma within Mental Health Misinformation}.}
    \item Do you justify how the proposed evaluation is sufficient and appropriate to the claims made? 
   \ethics{Yes, please refer to subsection “Methods” of RQ1. We justify the appropriateness of our LIWC and SAGE methods by referring to prior works that use them for similar tasks/objectives. For LLM-based approaches to detecting stigma and agreeement, we quantitatively validate the approaches by comparing machine-generated labels with human-generated labels (refer to Sections "Research Question 2 - Methods" and "Research Question 3 - Methods"). Further, we contextualize our findings on stigma and concurrence in response to health misinformation in the existing online misinformation literature.}
     \item Do you discuss what is ``the cost`` of misclassification and fault (in)tolerance?
    \ethics{Yes, we discussed the potential "cost" of misclassification and potential solutions to mitigate classification errors in the "Platform Design Recommendation" subsection, under "Targeted Interventions against Mental Health Misinformation".}
  
\end{enumerate}

\item Additionally, if you are using existing assets (e.g., code, data, models) or curating/releasing new assets...
\begin{enumerate}
  \item If your work uses existing assets, did you cite the creators?
    \ethics{Yes, we properly cite the manuscripts introducing the base LLMs we used for RQ1, RQ2 and RQ3}    
  \item Did you mention the license of the assets?
    \ethics{No, we did not due to space constraints. All open source LLMs are released under Apache 2.0, except for LLaMA which is released under its customized Apache-derived license and MentaLLaMA which is released under MIT license.}
  \item Did you include any new assets in the supplemental material or as a URL?
    \ethics{We will be releasing our mental health misinformation dataset and code used for the experiments in a public URL upon the publication of this manuscript.}
  \item Did you discuss whether and how consent was obtained from people whose data you're using/curating?
   \ethics{No, we did not due to space constraints. Given that we only collect publicly available data on Bitchute and Youtube, explicit content from people whose data we are curating is not required. We discuss steps to remove personally identifiable information from the dataset below.}
  \item Did you discuss whether the data you are using/curating contains personally identifiable information or offensive content?
   \ethics{No, we did not due to space constraints. Given that our collected dataset might contain personally identifiable information or offensive content, we follow best practices by only sharing video URLs and comment IDs in the public dataset. A fully hydrated dataset is available upon request, and evaluated on a case-by-case basis.}
\item If you are curating or releasing new datasets, did you discuss how you intend to make your datasets FAIR? %(see \citet{fair})?
 \ethics{No, we did not due to space constraints. We will be releasing our dataset on a publicly-facing Github repository (F, A) in .csv format (I). The dataset will be released under the Apache 2.0 license (R).}
\item If you are curating or releasing new datasets, did you create a Datasheet for the Dataset? % (see \citet{gebru2021datasheets})? 
 \ethics{Yes, we did. We will be releasing the Datasheet upon publication of this manuscript}
\end{enumerate}

\item Additionally, if you used crowdsourcing or conducted research with human subjects...
\begin{enumerate}
  \item Did you include the full text of instructions given to participants and screenshots?
    \red{NA}
  \item Did you describe any potential participant risks, with mentions of Institutional Review Board (IRB) approvals?
    \red{NA}
  \item Did you include the estimated hourly wage paid to participants and the total amount spent on participant compensation?
    \red{NA}
   \item Did you discuss how data is stored, shared, and deidentified?
    \red{NA}
\end{enumerate}
\end{enumerate}

% \appendix
% \appendixpage
\section{Appendix}
\setcounter{table}{0}
\renewcommand{\thetable}{A\arabic{table}}

\begin{table}[htbp!]
\small
\centering
\begin{tabular}{l@{}l@{}lll}
Platform\,\, & Category\,\,           & \# Comments & \# Commenters & Av. Length \\ \hline
YouTube\,\, & NMisinfo\,\, & 485738       & 283437
         &   102.9                \\
& Misinfo\,\,     & 84109        &    46694      &       97.6  \\
\hline
Bitchute\,\, & NMisinfo\,\, & 12747       & 4226         &   177                \\
& Misinfo\,\,     & 938        &    623      &       291  \\
\end{tabular}
\caption{Summary statistics for the comments associated with the labeled MHMisinfo and non-MHMisinfo videos for the \texttt{MentalMisinfo-Large} dataset} 
\label{tbl:summary_stat_comments_appendix}
\vspace{1cm}
\begin{tabular}{l@{}l@{}l@{}l@{}ll}
Platform & \,\,Metric  & \% Misinfo & \,\,\% NMisinfo\,\, & p-value & \(\chi^2\) \\ \hline
YouTube & \,\,Agreement\,\, & 11.05\%  & \,\,15.54\% & ***          & $3 * 10^4$                      \\
& \,\,Stigma    & 35.62\%        & \,\,22.49\%          &    ***           & 30057.51                      \\ \hline     
Bitchute & \,\,Agreement\,\, & 19.7\%  & \,\,11.2\% & ***        & 843.16                     \\
& \,\,Stigma     &   49.8\%   & \,\, 41.2\%        &    **           & 142.66                      \\ \hline   
\end{tabular}
\caption{Results summary for RQs 2 (Agreement Analysis) and 3 (Stigma Analysis) on YouTube and Bitchute comments for the \texttt{MentalMisinfo-Large} dataset}\label{tbl:rq23_comments_appendix}

\end{table}

\begin{table}[htbp!]
\centering
\small 
\setlength{\tabcolsep}{1.7pt}
\begin{tabular}{l|ll|l|l|l}
\textbf{LIWC}    & \textbf{NMisinfo (\%)} & \textbf{Misinfo (\%)} & \textbf{\% Diff} & \textbf{$q$-val}             & \textbf{$d$} \\ \hline
\multicolumn{6}{|c|}{\textbf{Affect}}                                                                                                                                              \\ \hline
Affect %(affect)            
& 7.17                         & 6.62                              & -7.72                   & *** (§)                       & 0.053             \\
         
& (6.5)                          & (6.2)                             & (-4.68)                  &                         & (0.029)               \\
Pos Emo %(posemo)  
& 4.62                        & 4.07                             & -11.87                 & *** (§)                       & 0.059               \\
& (2.87)                          & (2.63)                             & (-8.34)                  &                         & (0.031)               \\
Neg Emo %(negemo)  
& 2.51                         & 2.51                             & 0.26                  &                        & -0.001               \\
& (3.63)                          & (3.53)                             & (-2.56)                  &                        & (0.012)               \\
Anxiety %(anx)              
& 0.49                          & 0.26                              & -46.76                 & *** (§) & 0.102              \\
& (0.22)                          & (0.23)                             & (2.24)                  &                    & (-0.003)               \\
Sadness %(sad)              
& 0.65                         & 0.49                              & -24.87                   & *** (§) & 0.055               \\ 
& (0.37)                          & (0.32)                             & (-13.83)                  &                        & (0.02)               \\
\hline
\multicolumn{6}{|c|}{\textbf{Social}}                                                                                                                                              \\ \hline
social %(Social)           
& 9.44                           & 10.83                               & 14.81                    & *** (§)   & -0.123                \\
& (7.76)                          & (6.93)                             & (-10.75)                  &      ** (§)   & (0.09)               \\
family %(Family)            
& 0.98                          & 1.2                             & 24.89                 & *** (†)  & -0.057               \\
& (0.34)                          & (0.25)                             & (-28.13)                  &        * (§)         & (0.044)               \\
friend %(Friends)           
& 0.37                         & 0.46                             & 25.41                   &  *** (†)  & -0.038            \\
& (0.27)                          & (0.19)                             & (-30.07)                  &                       & (0.044)               \\
female %(Female)            
& 1.38                        & 0.4                            & -66.79                 & *** (§)                         & 0.214             \\
& (0.98)                          & (0.53)                             & ( -45.90)                  & *** (§)                       & (0.126)               \\
male %(Male)                
& 1.1                         & 3.7                           & 236.99                & *** (†) & -0.546              \\ 
& (1.07)                          & (0.67)                             & (-37.44)                  & *** (§)                      & (0.114)               \\
\hline
\multicolumn{6}{|c|}{\textbf{Somatic}}                                                                                                                                             \\ \hline
bio %(Biological Processes) 
& 2.68                        & 3.34                           & 24.77               & *** (†) & -0.106               \\
& (2.65)                          & (3.30)                             & (24.33)                  &       ** (§)     & (-0.1)               \\
body %(Body)                
& 0.6                         & 0.8                             & 31.72                  & *** (†)                        & -0.03             \\
& (1.06)                          & (0.72)                             & (-32.07)                  &       *** (§)                  & (0.092)               \\
health %(Health)            
& 1.13                         & 1.08                           & -3.90               & ** (§) & 0.0117               \\
& (0.7)                          & (1.56)                             & (122.92)                  & *** (†)                        & (-0.269)               \\
sexual %(Sexual)           
& 0.11                        & 0.16                             & 40.00                & *** (†) & -0.03              \\
& (0.78)                          & (0.69)                             & (-11.69)                  &                     & (0.025)               \\
ingest %(Ingest)            
& 0.53                     & 1.06                             & 98.35                & *** (†) & -0.157 (†)               \\ 
& (0.33)                          & (0.35)                             & (6.83)                  &                       & (-0.011)               \\
\hline
\multicolumn{6}{|c|}{\textbf{Drives}}                                                                                                                                              \\ \hline
drives %(Drives)           
& 5.11                     & 5.99                            & 17.18                & *** (†) & -0.104             \\
& (4.87)                          & (5.18)                             & (6.46)                  &                         & (-0.04)               \\
affiliation %(Affiliation)  
& 1.76                      & 2.09                           & 19.14               & *** (†)                         & -0.063            \\
& (1.0)                          & (0.96)                             & (-4.54)                  &                      & (0.014)               \\
achieve %(Achievement)     
& 0.82                         & 0.85                             & 2.64                  &  & -0.006             \\
& (0.72)                          & (0.63)                             & (-11.8)                  &                        & (0.032)               \\
power %(Power)              
& 1.42                       & 1.84                           & 29.21                  & *** (†) & -0.097            \\
& (2.19)                          & (2.25)                             & (2.64)                  &                        & (-0.011)               \\
reward %(Reward)            
& 1.15                        & 1.27                           & 10.72                   & *** (†)                        & -0.033            \\
& (0.91)                          & (1.06)                             & (16.21)                  &                        & (-0.051)               \\
risk %(Risk)                
& 0.44                         & 0.46                          & 4.824               & * (†) & -0.0099             \\ 
& (0.49)                          & (0.60)                             & (23.6)                  &                       & (-0.053)               \\
\hline
\multicolumn{6}{|c|}{\textbf{Personal Concerns}}                                                                                                                                  \\ \hline
work %(Work)                
& 0.86                         & 0.98                             & 14.68               & *** (†) & -0.037              \\
& (1.38)                          & (1.68)                             & (21.7)                  &    ** (§) & (-0.088)               \\
leisure %(Leisure)          
& 0.72                     & 0.79                             & 9.39                 & *** (†) & -0.16              \\
& (0.76)                          & (0.99)                             & (30.25)                  &  * (†)   & (-0.08)               \\
home %(Home)               
& 0.2                        & 0.14                             & -31.41                 & *** (§) & 0.042             \\
& (0.18)                          & (0.17)                             & (-4.74)                  &                         & (0.005)               \\
money %(Money)             
& 0.17                       & 0.21                           & 24.39              & *** (†)                        & -0.029            \\
& (3.47)                          & (4.09)                             & (18.12)                  &                       & (-0.035)               \\
relig %(Religion)           
& 0.45                       & 0.88                            & 95.96                & *** (†) &  -0.112            \\
& (0.53)                          & (0.55)                             & (3.55)                  &                        & (-0.006)               \\
death %(Death)             
& 0.11                          & 0.33                            & 187.49                 & *** (†) & -0.154     \\ 
& (0.4)                          & (0.54)                             & (33.46)                  &                         & (-0.055)               \\ \hline         
\end{tabular}
\caption{Relevant LIWC categories in the comments, along with results of Student's $t$-tests for the \texttt{\texttt{MentalMisinfo-Large}}. Rows with section marks (§) following q-value indicates that usage among non-MHMisinfo (NMisinfo) comment is significantly higher. Rows with daggers (†) following q-value indicates that usage among MHMisinfo (Misinfo) comment is significantly higher. Bitchute-related metrics are presented with round brackets. * indicates \(q < 0.05\), ** indicates \(q < 0.01\), *** indicates \(q < 0.001\)} \label{tbl:liwc_appendix}
\end{table}

\begin{table*}
\centering
\small
\begin{tabular}{llllllll}
Platform & Category        & \# Videos & \# Creators & Av. Length (s) & Av. \#Views & Av. \#Likes & Av. \#Comments \\ \hline
YouTube & NMisinfo & 6308  & 4259 & 28.77                & 75008        & 3186          & 68              \\
& Misinfo     & 986        & 875          & 35.46               & 25556         & 1569           & 39  \\ \hline     
Bitchute & NMisinfo & 741       & 318         &     109           & 795         & 20           & 17              \\
& Misinfo     & 146        & 116          &    141           & 1503          & 11            & 6  \\ \hline   
\end{tabular}
\caption{Summary statistics for the labeled MHMisinfo (Misinfo) and non-Misinfo (NMisinfo) videos for the \texttt{MentalMisinfo-Large} dataset}\label{tbl:summary_stat_videos_appendix}
\vspace{1cm}

\begin{tabular}{lllp{8cm}}
\textbf{Keyword}              & \textbf{\# Videos (Youtube)} & \textbf{\# Videos (Bitchute)} & \textbf{Example Video Title}                                                                              \\ \hline
ocd                           & 1938                  & 163                & If You Said Your Intrusive/OCD Thoughts Out Loud \#shorts                                                 \\
bipolar                       & 1913                  & 197                & Bipolar Pass? Am I the a hole?                                                                            \\
depression                    & 4102                  & 490                & Black Panther Star Letitia Wright Says God Saved Her from Depression |                         \\
borderlinepersonalitydisorder & 478                   & 5                & What is BPD or Borderline Personality Disorder? More on my page!           \\
ptsd                          & 2309                  & 480                & 100\% VA Rating for Mental Disorders - PTSD, Anxiety, Depression, ...        \\
schizophrenia                 & 980                   & 501               & Schizophrenia or DEMONS!?                                                                                 \\
mentalillness                 & 901                   & 435               & Ben Shapiro discusses Why he thinks Gender Dysphoria is a Mental illness!                     \\
anorexia                      & 524                   & 38                & Do you know What anorexia really feels like?                                                              \\
psychosis                     & 522                   & 490                & NEUROSCIENTIST WEED Causes PSYCHOSIS and SCHIZOPHRENIA       \\
autism                        & 4166                  & 508                & There is no cure for Autism, we just manage the day.                                             \\
% bpd                           & 545                   & 0.077                & TBD                                                                                                       \\
mentaldisorder                & 629                   & 27                & \#pov : mental disorder more people need to know about  \\
self-harm                     & 120                   & 499                & Y/n's suicidal and Pansy makes fun of herTW: mention of self harm                                         \\
panic-attack                  & 667                   & 4                & Panic Attack Right Before I Walk Down The Aisle                                                   \\
obsessivecompulsivedisorder   & 562                   & 1                & Queen Elizabeth's crippling obsessive compulsive disorder                                       \\
suicidal                      & 1255                  & 484                 & ``Why Gender Dysphoric People are feeling Suicidal" - Jordan Peterson                              \\
schizophrenic                 & 592                   & 196               & PARANOID SCHIZOPHRENIC CRAZY STORY                                                    \\
adderall                      & 232                   & 72                & ADDERALL SHORTAGE: Here's What You Can Do About It                           \\
anxiety                       & 4009                  & 499                & My Son Has Separation Anxiety!!!                                                                \\
antidepressants               & 545                   & 161                 & Joe Rogan Opens Up The Truth About Antidepressants              \\
eating disorder               & 1987                  & 28                & Un-glamorizing Eating Disorder Recovery                                                                   \\
adhd                          & 2462                  & 489                & How To Discuss ADHD With Difficult Teachers                                      \\
mentalhealth                  & 3973                  & 493                & Reflections: Jordan Neely \& Mental Health \#shorts       \\ \hline                                                 
\end{tabular}
\caption{Keywords used in the data collection process, number of videos gathered per keyword for each platform and an examplar video title with the keyword} \label{tbl:dist_keywords_appendix}
\vspace{1cm}

\end{table*}

\begin{table}[htbp!]
\centering
\setlength{\tabcolsep}{2pt}
\begin{tabular}{c|lrlr}
\textbf{Platform} & \multicolumn{2}{l|}{\textbf{Misinfo}}           & \multicolumn{2}{l|}{\textbf{NMisinfo}} \\ \hline 
& $n$-gram & SAGE & $n$-gram        & SAGE       \\ \hline
& bread              & 1.931                    & mikko          & 0.133                                                           \\
& brother                       & 1.452                    & alice          & 0.132                              \\
& evil                        & 1.437                    & stim           & 0.131                             \\
& schizophrenia                        & 1.372                    & birthday           & 0.128                             \\
& kill                        & 1.136                    & thomas           & 0.127                              \\
& bipolar                      & 1.135                    & hair         & 0.125                             \\
& crazy                  & 1.039                    & adhd          & 0.123                             \\
& meds                       & 1.035                   & mens           & 0.122                              \\
& mentally                        & 0.985                    & precious           & 0.116                              \\
& free                        & 0.889                    & relatable           & 0.112                              \\
& illness                 & 0.795                   & relate           & 0.108                            \\
\textbf{YouTube} & speaking                   & 0.794                   & beautiful           & 0.105                            \\
& he’s                        & 0.788                    & movement           & 0.105                            \\
& pray                       & 0.771                    & coping          & 0.104                              \\
& control                        & 0.755                    & baby           & 0.101                              \\
& excuse                        & 0.735                   & anxiety           & 0.1                              \\
& jesus                     & 0.706                    & adorable          & 0.098                             \\
& looks                     & 0.703                    & toxic          & 0.098                             \\
& taking                        & 0.663                    & teacher           & 0.0976                              \\
& society                       & 0.654                    & daughter         & 0.0956                              \\
& dude                      & 0.643                    & class           & 0.0936                              \\
& drugs                       & 0.619                    & parents          & 0.0926                              \\
& respect                       & 0.617                    & spectrum           & 0.0914                              \\
& book                       & 0.590                    & autism           & 0.0904                             \\
& sounds                        & 0.590                    & school           & 0.0876                               \\ \hline                       
& sterilize              & 2.442                    & like           &  \(3.7 * 10^{-4}\)                                                            \\
& vaccines                       & 1.722                    & women           &   \(3.6 * 10^{-4}\)                                \\
& medical                        & 1.698                    & year           &  \(3.2 * 10^{-4}\)                              \\
& drugs                       & 1.462                    & back            &  \(2.9 * 10^{-4}\)                              \\
& doctors                      & 1.451                    & would            &  \(2.4 * 10^{-4}\)                              \\
& three                     &  1.139                    & time         &  \(2.3 * 10^{-4}\)                             \\
&  vaccine                 & 1.109                    & woman            &  \(1.97 * 10^{-4}\)                              \\
& later                        & 1.005                    & thats            &  \(1.93 * 10^{-4}\)                              \\
& gender                       & 0.989                    & look            &  \(1.84 * 10^{-4}\)                              \\
& shall                       & 0.970                    & thing           & \(1.61 * 10^{-4}\)                                \\
& trust                 & 0.945                   & want           & \(1.5 * 10^{-4}\)                              \\
\textbf{Bitchute} & covid                   & 0.935                    & dont            & \(1.34 * 10^{-4}\)                             \\
& brain                        & 0.902                    & show           & \(1.33 * 10^{-4}\)                              \\
& birth                      & 0.877                    & fuck           & \(1.3 * 10^{-4}\)                                \\
& bitchute                      & 0.826                    & right             & \(1.24 * 10^{-4}\)                                \\
& channel                       & 0.825                    & police           & \(1.18 * 10^{-4}\)                           \\
&  found                    & 0.824                    & someone          & \(1.14 * 10^{-4}\)                               \\
& schools                    & 0.817                    & need          & \(1.12 * 10^{-4}\)                               \\
&  fear                       & 0.816                    & doubt           & \(9.64 * 10^{-5}\)                              \\
& started                        & 0.751                    & folks            & \(9.512 * 10^{-5}\)                             \\
& list                     & 0.742                    & going           & \(9.346 * 10^{-5}\)                                \\
& health                       & 0.734                    & cant          & \(9.022 * 10^{-5}\)                              \\
& body                      & 0.723                    & mental            & \(8.904 * 10^{-5}\)                                \\
& also                       & 0.689                    & cops           & \(8.044 * 10^{-5}\)                               \\
& month                       & 0.687                    & good           & \(7.9 * 10^{-5}\)                              \\ 
\hline
\end{tabular}
\caption{Top-25 most discriminative n-grams for MHMisinfo (Misinfo) comments and non-MHMisinfo (NMisinfo) comments for the \texttt{MentalMisinfo-Large} dataset, identified through the SAGE algorithm.} \label{tbl:sage_appendix}
\end{table}

\begin{table}
\centering\small
\begin{tabular}{|p{8cm}|} \hline 
    \textit{You are an expert psychiatrist who has comprehensive knowledge on all mental health conditions and misinformation surrounding it. You are helpful, so you will try your best to give accurate answers. Now, thoroughly look at the following examples which contains the audio transcription and rationale on whether a video contains misinformation regarding mental health or not.} \\
    \textit{Example: [Example 1 Data].} \textit{Answer: [Answer]} \\
    ...... \\
    \textit{Example: [Example 5 Data].} \textit{Answer: [Answer]} \\
    \textit{Now, only using the video data provided below, answer whether the video contains mental health misinformation or not. You must include the word ``yes" or ``no" within your answer, then give your reasoning.} \\
    \textit{[Target Data].} \textit{Answer:} \\
\hline 
\end{tabular}
\caption{Prompt input template for detecting MHMisinfo using LLMs} \label{tbl:prompt_misinfo}
\vspace{1cm}
\begin{tabular}{|p{8cm}|} \hline 
\textit{You are an expert in linguistics and social media data. Please analyze the following examples where we determine whether a comment agrees with a video on mental health based on the video's transcription. There are five examples, one on each line. Each example contains the comment and the human-generated answer:} \\
 \textit{Example 1: Video Transcript: [Example 1 Data] Comment: [Comment] Answer: [Yes/No + Reasoning]} \\
 ......\\
\textit{Example 5: Video Transcript: [Example 5 Data] Comment: [Comment] Answer: [Yes/No + Reasoning]} \\
 \textit{Now, given what you learned from the five examples above, please determine whether the comment agrees with the video or not. Answer with yes or no, and then give your reasoning.}\\
\textit{Video Transcript: [Target Data] Comment: [Comment] Answer: } \\
\hline 
\end{tabular}
\caption{Prompt input template for agreement assessment using LLMs %in comments
}\label{tbl:prompt_stigma_appendix}
\vspace{1cm}
\begin{tabular}{|p{8cm}|} \hline 
\textit{You are an expert in psychiatry and social media data. Please analyze the following examples where we determine whether a YouTube comment actively generates stigma towards people with mental health conditions. A comment is not stigmatizing if it only talks about one's personal experience with mental health. There are five examples, one on each line. Each example contains the comment and the human-generated answer:} \\
 \textit{Example 1: Comment: [Comment] Answer: [Yes/No + Reasoning]} \\
 ......\\
 \textit{Example 5: Comment: [Comment] Answer: [Yes/No + Reasoning]} \\
 \textit{Now, given what you learned from the five examples above, please determine whether the comment actively generates stigma against people with mental health conditions. Answer with yes or no, and then give your reasoning. Answer no if the comment's author only talks about their experience: }\\
 \textit{Comment: [Comment] Answer: } \\
\hline 
\end{tabular}
\caption{Prompt input template for stigma assessment using LLMs %in comments
}\label{tbl:prompt_agreement_appendix}
\end{table}

\end{document}